\documentclass{aa}

\newcommand{\hubert}[1]{\textcolor{black}{#1}} % Hubert Klahr
\usepackage{graphicx}
\usepackage{txfonts}
\usepackage{float}
\usepackage{dblfloatfix}
\usepackage{natbib}
\bibpunct{(}{)}{;}{a}{}{,} % to follow the A&A style
\usepackage{hyperref}
\hypersetup{colorlinks=true, urlcolor=blue, linkcolor=blue, citecolor=blue}
\usepackage{wrapfig}
\usepackage{comment}
\usepackage{svg}

\title{The impact of pebble flux regulated planetesimal formation on giant planet formation}
\titlerunning{The impact of pebble flux regulated planetesimal formation on giant planet formation}

\author{
    Oliver Voelkel\inst{\ref{mpia}}
    \and
    Hubert Klahr\inst{\ref{mpia}}
    \and
    Christoph Mordasini\inst{\ref{unibe}}
    \and
    Alexandre Emsenhuber\inst{\ref{lpl},\ref{unibe}}
    \and
    Christian Lenz\inst{\ref{mpia}}
}
\authorrunning{O. Voelkel et al.}

\institute{
    Max Planck Institute for Astronomy, Heidelberg, Königstuhl 17, 69117 Heidelberg, Germany \\
    \email{voelkel@mpia.de} \label{mpia}
    \and
    Physikalisches Institut, University of Bern, Gesellschaftsstrasse 6 CH 3012 Bern, Switzerland \label{unibe}
    \and
    Lunar and Planetary Laboratory, University of Arizona, 1629 E. University Blvd., Tucson, AZ 85721, USA \label{lpl}
}
    
\date{Received 3 April 2020 / Accepted DD MMM YYYY}

\abstract
   {
    Forming gas giant planets by the accretion of 100$\,$km diameter planetesimals, a typical size that results from self-gravity assisted planetesimal formation, is often thought to be inefficient. Many models therefore use small km-sized planetesimals, or invoke the accretion of pebbles. 
    Furthermore, models based on planetesimal accretion often use the ad hoc assumption of planetesimals distributed radially in a minimum mass solar nebula fashion.
   }
   {
    We wish to investigate the impact of \hubert{various initial radial} density distributions \hubert{in planetesimals with a dynamical model for the} formation of planetesimals on the \hubert{resulting} population of planets. In doing so, we highlight the directive \hubert{role} of the early stages of \hubert{dust evolution into pebbles and planetesimals in} the circumstellar disk on the following planetary formation.
   }
   {
   We have implemented a two population model for solid evolution and a pebble flux regulated model for planetesimal formation into our global model for planet population synthesis. This framework is used to study the global effect of planetesimal formation on planet formation. As reference, we compare \hubert{our} dynamically formed planetesimal surface densities with \hubert{ad-hoc} set distributions of different \hubert{radial} density slopes \hubert{of planetesimals}.
   }
   {
    Even though required, it is not solely the total planetesimal disk mass, but the planetesimal surface density slope and subsequently the formation mechanism \hubert{of planetesimals}, that \hubert{enables} planetary growth via planetesimal accretion. Highly condensed regions of only 100$\,$km sized planetesimals in the inner regions of circumstellar disks can lead to gas giant growth. 
   }
   {
   Pebble flux regulated planetesimal formation strongly boosts planet formation \hubert{even if the planetesimals to be accreted are 100$\,$km in size, because it is a highly effective mechanism to create a steep planetesimal density profile}. \hubert{We find this to} lead to the formation of giant planets inside 1$\,$au by 100$\,$km \hubert{already} by pure planetesimal accretion. \hubert{Eventually adding also pebble flux regulated pebble accretion and planetesimal based embryo formation will further complement this picture.}
   }

\keywords{   
   	planetesimal formation --
    planetesimal accretion --
    pebble accretion --
    population synthesis
}

\begin{document}
\maketitle
%
%________________________________________________________________
%
\section{Introduction}

\subsection{Physical background}

A current conundrum of planetesimal accretion in the core accretion scenario of planet formation is that for 100$\,$km planetesimals it \hubert{appears to require} an unreasonably high \hubert{disk} mass to be an effective mechanism for giant planet formation within the lifetime of a circumstellar disk \hubert{\citep{Fortier2013}}. The accretion of smaller objects with a higher effective cross section, like either km-sized planetesimals \citep{ida2004toward} or cm sized bodies known as pebble accretion \citep{ormel2010effect} is often described as the solution for giant planet formation and has been studied widely by \cite{KlahrBodenheimer2006}, \cite{lambrechts2012rapid}, \cite{levison2015growing} and \cite{bitsch2015growth} to name just a few. While we restrain to make a statement on the efficiency of pebble accretion, the scenario of a planetary core accreting inward drifting pebbles \hubert{also} lacks an explanation on how, where and when this planetary core first forms. \hubert{Planetesimals are typically too small for efficient pebble-accretion \citep{ormel2010effect}, thus a pebble accreting embryo could well have formed from planetesimal collisions.} This crucial step adds room to discuss the formation of planetesimals and subsequently their role in planetary core and planet formation.
From \cite{tanaka1999growth} we know that the accretion rate of planetesimals depends on the planetesimal size and linearly on the planetesimal surface density. Constraining their size is an active field of research. While some studies infer \hubert{that} the current size of asteroid belt objects is well constrained and found to be in the order of magnitude of 100$\,$km in diameter \citep{bottke2005linking,walsh2017identification,Delbo2017}, other studies find that the size distribution found today merely reflects which sizes are most resilient to clearing, and therefore suggest a smaller primordial size \citep{zheng2017planetesimal}. The observed size distribution could also arise from the growth of planetesimals of originally 100$\,$m in size \citep{weidenschilling2011initial}. In the Kuiper belt, the size distribution has a similar shape as predicted by simulations including the \hubert{streaming} instability between 10 and 100$\,$km \citep{schafer2017initial}, indicating large initial sizes. On the other hand, recent discoveries of Kuiper belt objects via stellar occultations rather indicate a size of 1-2$\,$km \citep{arimatsu2019kilometre}. Small initial sizes of 0.4 - 4$\,$km are also inferred theoretically by \cite{schlichting2013initial}.
Also, the surface density profile of planetesimals for extrasolar systems is unknown. Studies of our own solar system motivated the minimum mass solar nebula (mmsn) hypothesis \citep{weidenschilling1977distribution,hayashi} that results in a power law drop of the planetesimal surface density with a decay of $\Sigma_P \propto r^{-1.5}$. Observations of solid material in disks \citep{Andrews_2010} and the widely used $\alpha$-disk model for the viscous evolution of an accretion disk \citep{shakura1973black} suggest a shallower density distribution of $\Sigma_P \propto r^{-0.9}$ \hubert{for radially constant $\alpha$}. The observed solid material however is not planetesimals, but the dust in the circumstellar disk, as the distribution of planetesimals in protoplanetary disks is currently unobservable.  \cite{Lenz_2019} model the formation of planetesimals based on the solid evolution of a viscously evolving disk, assuming that planetesimals form proportional to the \hubert{time-dependent local} radial pebble flux. They find that the profile of the planetesimal surface density becomes significantly steeper ($\Sigma_P \propto r^{-2.1}$) than the initial dust, pebble \hubert{and} gas density ($\Sigma \propto r^{-0.9}$). 
This mass transfer results in an increase in the planetesimal surface density in the inner circumstellar disk by orders of magnitude without increasing the total mass in planetesimals. Since the accretion rate of planetesimals is proportional to the local planetesimal surface density, these highly condensed planetesimal zones are promising to have a drastic effect on planetary growth.
\subsection{Previous models}
Before discussing some of the previous work, we would like to distinguish between a global planet formation model and a model for planet population synthesis. While a model for planet population synthesis contains (or should contain) a global formation model, this does not yield vice versa. Key to the population synthesis approach is that the model is complex enough to take into account the physical effects that are deemed crucial for planet formation, yet its single system computational cost is low enough so that it can be used to study a wide range of parameters. Only this will enable a statistical comparison with observational data. For this purpose, it is vital to find ways to simplify complex physical processes and merge them to a more complex framework, without loosing the essence of their nature. 
The formation of planetesimals is such a process and the one dimensional formation model by \cite{Lenz_2019} is such an attempt.
Previous work on the accretion of planetesimals for planetary growth like \cite{johansen2019exploring}, \cite{mordasini2018planetary} or \cite{ida2004toward} all use initial distributions of planetesimals and initially placed planetary embryos, while neglecting the presence of pebbles. Other formation models like \cite{bitsch2015growth} or \citet{brugger2018metallicity} model planetary growth by the accretion of pebbles and initially set planetary embryos, while neglecting the formation, or accretion of planetesimals. Yet, a model that contains both pebble and planetesimal accretion, while also taking the formation of planetesimals and planetary embryos into account is still pending.

We have chosen to thus improve our planet population synthesis model by a \hubert{"disk consistent"\footnote{"disk consistent" means that both dust evolution and planet formation use the same disk model, including viscosity, density and temperature evolution.}} model for solid evolution \citep{birnstiel2012simple} and planetesimal formation \citep{Lenz_2019} to take the early stages of the disks evolution into account. This early phase determines the planetesimal surface density distribution, the radial pebble flux evolution, the formation of planetary cores and therefore planet formation as a whole. For our study, we will focus on the formation and accretion of planetesimals. 
We will display the impact of the planetesimal surface density and its formation on the population of planets. We will show that the accretion by 100$\,$km sized planetesimals is in fact a highly efficient growth mechanism for planets, due to the highly condensed planetesimal regions in the disk.
Furthermore we will give an overview over the future possibilities that arise from our newly implemented modules. 
This paper is outlined as followed: in Sect.~\ref{planetesimal_formation_model} the planetesimal formation model is explained, as well as the newly implemented solid evolution model on which it is based on. Sect.~\ref{Bern_model_for_population_synthesis} will give insight into the population synthesis framework and how it was modified for our purpose. The changes in $\Sigma_P$ in the population synthesis code, as well as the newly computed synthetic populations are presented in Sect.~\ref{Results}. Sect.~\ref{Sec:Discussion} will discuss the results followed by a brief summary and an outlook on our new possibilities and future work in Sect.~\ref{summary}.
\section{The planetesimal formation model}
\label{planetesimal_formation_model}
\subsection{The two population solid evolution model}
The two population model for solid evolution by \cite{birnstiel2012simple} is a parameterized approach to model the evolution and growth of dust and cm sized bodies in circumstellar disks. A detailed description of the model can be found in \cite{birnstiel2012simple} and \cite{Lenz_2019}. This chapter gives a brief outline of the assumptions and displays the most important reasons why we have chosen to use it in our framework. Our goal is to implement a fast computing, one dimensional, parameterized algorithm for solid evolution that is well tested and in good agreement with more sophisticated models. Key of the performance of the two population approximation is a parameterized mass ratio $f_m(r)$ as a function of orbital distance $r$ between two populations of solids, that depends on whether the growth of the particles is limited by drift or by fragmentation. Each time step, the model solves one advection-diffusion equation given by
\begin{align}
    \frac{\partial \Sigma_s}{\partial t} + 
    \frac{1}{r} \frac{\partial}{\partial r} 
    \left[ 
    r \left(
    \Sigma_s \bar{u} - D_{g} \Sigma_g \frac{\partial}{\partial r}
    \left(
    \frac{\Sigma_s}{\Sigma_g}
    \right)
    \right)
    \right]
    = 0
    \label{advection_diffusion_equation}
\end{align}
with $\Sigma_s$ as the total solid surface density without planetesimals, $\Sigma_{g}$ as the gas surface density and $D_{g}$ as the gas diffusion coefficient and $t$ and $r$ as time and radial distance.  $\bar{u}$ describes the weighted velocity of the total solid density and is defined as 
\begin{align}
    \bar{u} = (1-f_m(r)) \cdot u_0 + f_m(r) \cdot u_1
    \label{weighted_velocity}
\end{align}
where $f_m$ is the fit parameter for the mass ratio between the two populations. $u_0$ and $u_1$ describe their velocities, while the surface densities of the two populations are given as
\begin{align}
    \Sigma_0 (r) &= \Sigma_s (r) \cdot (1-f_m(r)) 
    \label{twopop_dust_density}
    \\
    \Sigma_1 (r) &= \Sigma_s(r) \cdot f_m(r)
    \label{twopop_pebble_density}
\end{align}
The two populations are defined by their Stokes number. Particles with a small Stokes number of \hubert{$\text{St} << 1$} are \hubert{strictly} coupled to the evolution of the gas, whereas particles with \hubert{$\text{St} \geq 1$} are not. $\Sigma_0$ describes the smaller population, that can be seen as dust, \hubert{subject to diffusion and transport with the gas,} while $\Sigma_1$ describes the larger population, that can be seen as pebbles \hubert{, which on top of being diffused by the gas are also sedimenting towards the midplane and drifting towards pressure maxima, for instance towards the star}. The fit parameter $f_m$ has been derived by comparing the two population model to the more sophisticated dust model from \cite{Birnstiel_2010}. The values for $f_m$ that were the best fit are given as 
\begin{align}
    f_m = \left\{\begin{array}{ll} 
        0.97, & \text{drift-limited case} \\
        0.75, & \text{fragmentation-limited case}
        \end{array}\right..
        \label{f_m_parameter}
\end{align}
These are also the ones that we used in our simulations.
\hubert{One can see the effect of this implementation in Fig.\ \ref{fig:disk_evolution}, where the ratio between dust and pebbles varies with space and time, visible in the two blue curves.}
\subsection{Pebble flux regulated planetesimal formation}
The full model and its results are described in \cite{Lenz_2019} in greater detail, we thus outline here only the basic physical assumptions behind this one dimensional approach and summarize the most important equations and results. The principle behind this parameterized model is that planetesimals form \hubert{by a local continuous mechanism that converts a certain fraction of the pebbles drifting by into planetesimals. Thus in principle it acknowledges that pebbles want to drift inward and that one can form more planetesimals if more material comes by. Many different planetesimal formation prescriptions can therefore be parameterised in such a fashion. Be it in the frame work of turbulent clustering \citep{Cuzzi2010,Hartlep2020} or streaming instabilities \citep{Johansen2009,schafer2017initial} or local trapping in zonal flows \citep{Johansen2007, Johansen2011, Dittrich2012, Drazkowska2017} or in vortices \citep{Raettig2014, Raettig2018}, the formation is always limited by how much a region receives in fresh pebbles, after consuming the locally available ones. Thus our parameterisation is per se model independent. Different scenarios might lead to the same conversion rates for the pebble flux. The parameters we need is the fraction $\epsilon$ of pebbles that is converted into planetesimals after having drifted over a distance of $d$ within the disk. }

\begin{figure}
\centering
\includegraphics[width=0.45\textwidth]{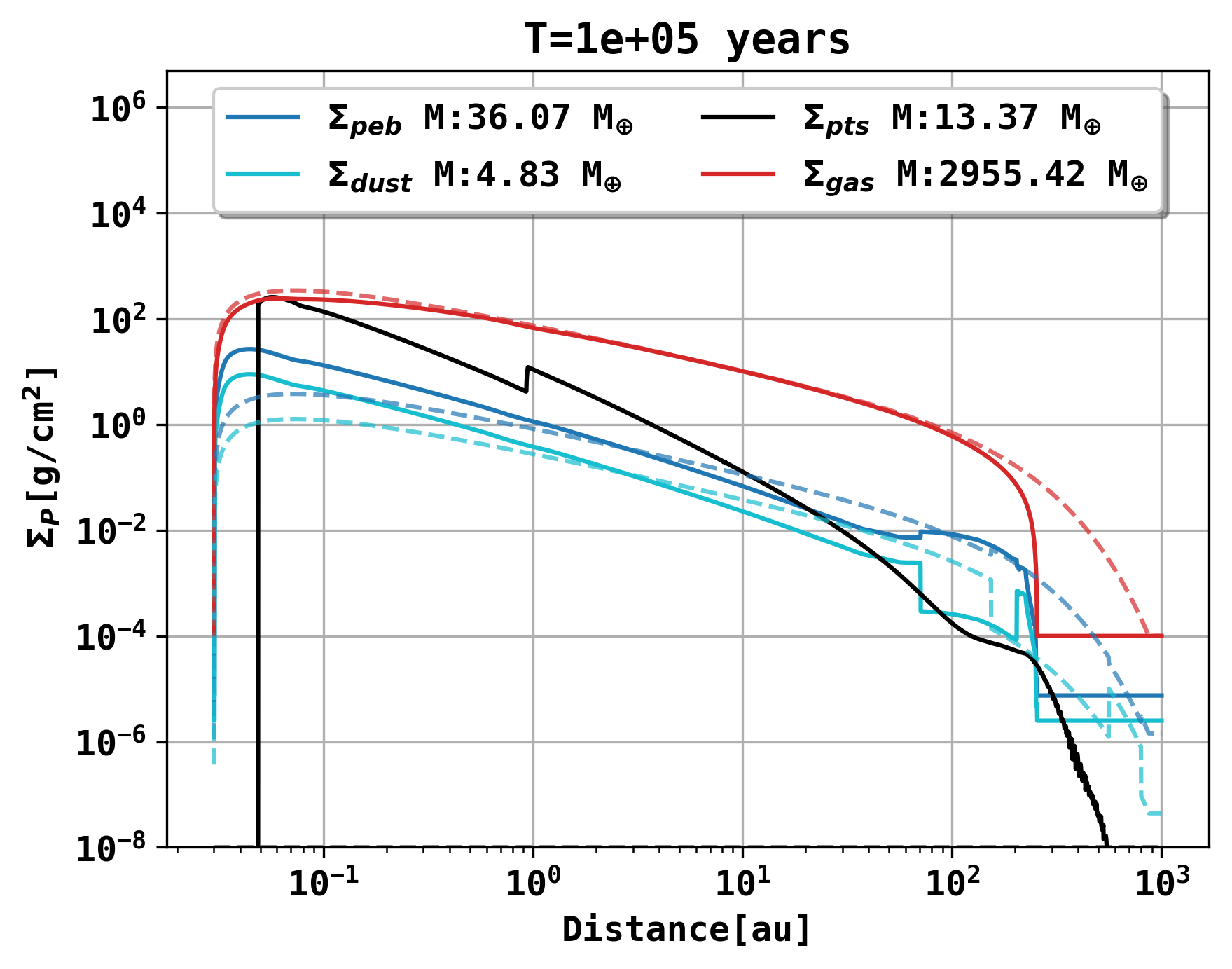}
\includegraphics[width=0.45\textwidth]{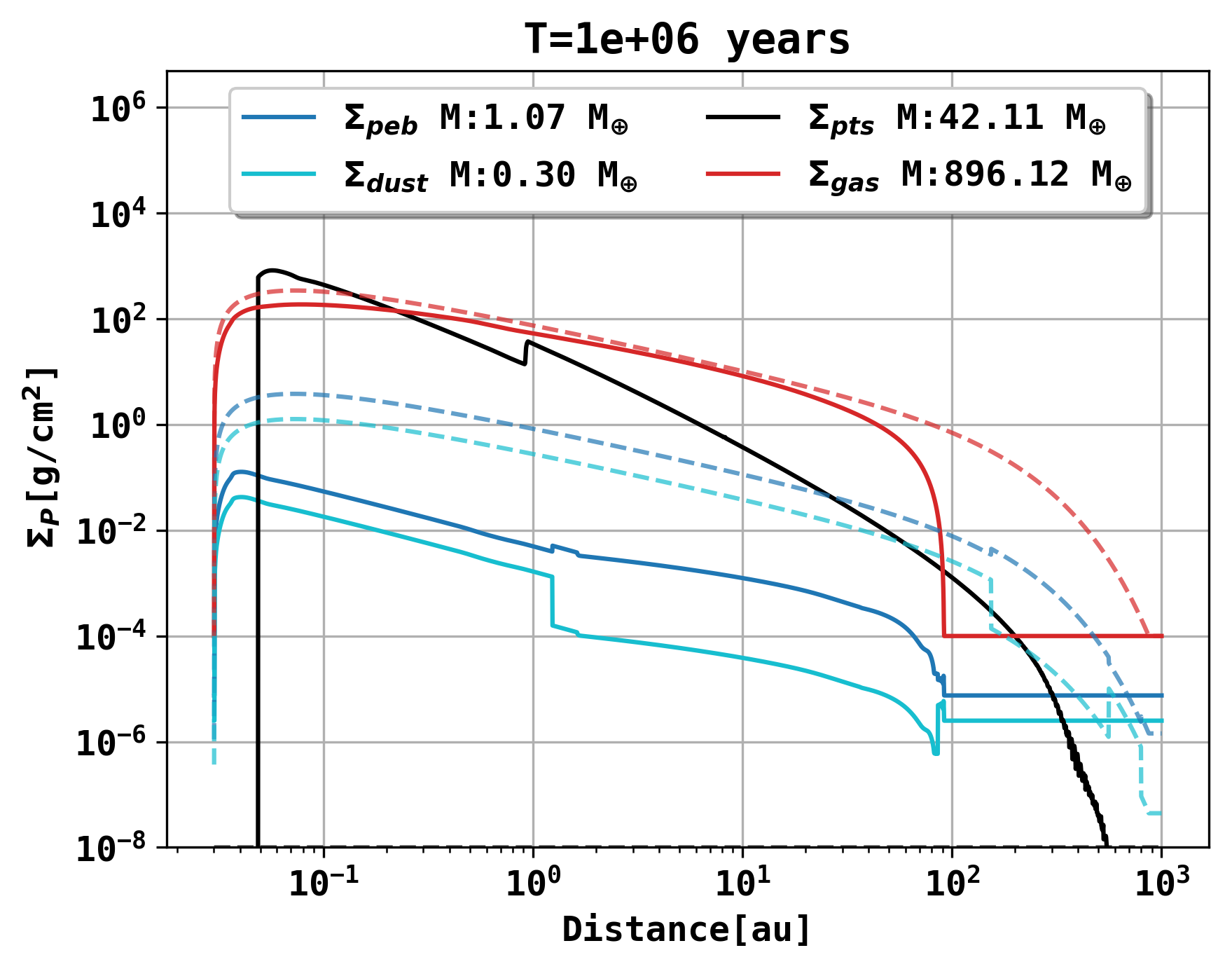}
\includegraphics[width=0.45\textwidth]{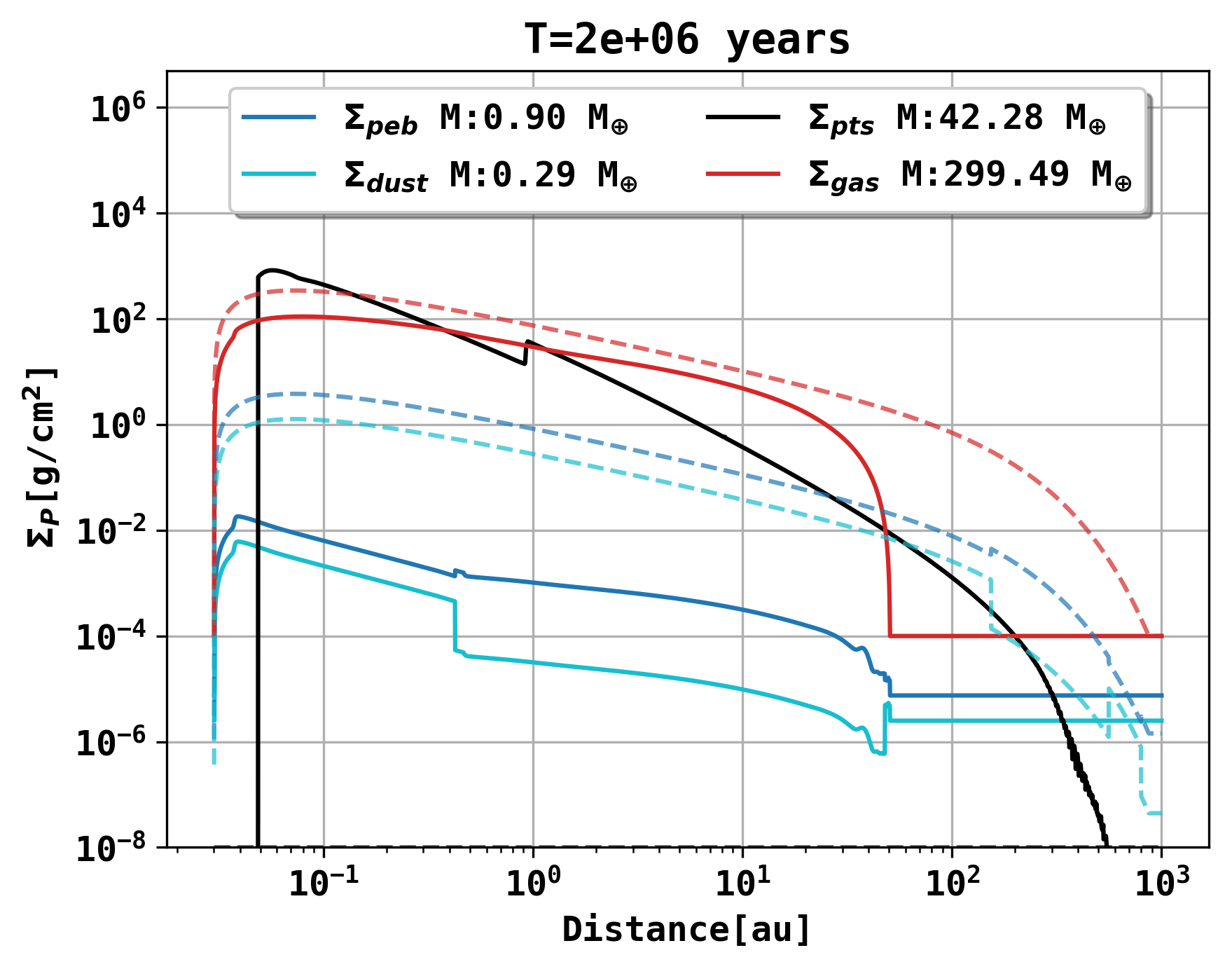}
\caption{\hubert{Exemplary} disk evolution \hubert{including our dynamical model for planetesimal formation} after 0.1$\,$Myrs, 1$\,$Myrs and 2$\,$Myrs. We show the surface density for the dust, pebbles, planetesimals, gas and their individual disk masses. The dashed lines refer to the initial profile of the corresponding density. This run does not contain a planetary embryo, it only evolves the disk dynamically. The total disk gas mass is given as 0.012$\,$M$_{\odot}$ with a dust to gas ration of 1.5$\%$ and $\alpha = 10^{-3}$. The exponential cutoff radius of the disk is at 137$\,$au, the inner radius at 0.03$\,$au and the evaporation rate is given as $2.87\times10^{-5}$ $\,$M$_{\odot}$/year. The planetesimal and solid evolution parameters can be found in table ~\ref{tab:initial_parameters}. \hubert{Note the effect of the iceline visible in the kink in the planetesimal distribution around $1\,$au and the effect of drift vs. fragmentation limited pebble size in the radially varying dust to pebble ratio.}}
\label{fig:disk_evolution}
\end{figure}

\hubert{
We can motivate these parameters easily in our paradigm of trapping zones that are slowly evolving coherent flow structures in protoplanetary disks like vortices and zonal flows \citep{2018haex.bookE.138K}, which can form everywhere, which live only for a limited time and thus only trap a fraction of drifting pebbles.
In these traps, pebbles get sufficiently concentrated, that regulated by streaming and Kelvin Helmholtz instabilities, planetesimal formation will be triggered. 
The planetesimal formation rate is generally proportional to the radial pebble flux}
\begin{align}
  \dot{M}_{\text{peb}}   := 2 \pi r \sum_{
  \text{St}_\text{min} \leq \text{St} \leq \text{St}_\text{max} 
  } 
  {
  | v_\text{drift} (r,\text{St}) | \Sigma_\text{s} (r,\text{St} )
  }
  \label{pebble_flux}
\end{align}
with $v_\text{drift}$ as the drifting velocity of the particles and $\text{St}_\text{min}$ and $\text{St}_\text{max}$ as the minimal and maximal Stokes number for which a particle is considered a pebble. $v_\text{drift}$ is given as 
\begin{align}
v_\text{drift}(r,\text{St})  =  \frac{ \text{St} }{\text{St}^2 + 1} \frac{h_g(r)}{r} \frac{\partial ln P(r)}{\partial ln r} c_s(r)
\end{align}
with $P(r)$ as gas pressure, $h_g(r)$ as gas pressure scale height ($h_g(r) = c_s(r) / \Omega(r)$) and $c_s(r)$ as sound speed. $\Omega(r)$ is given as the orbital frequency at the radial distance $r$. 
The \hubert{source term for planetesimals, i.e. for} $\Sigma_P$ is then given as \citep{Lenz_2019}
\begin{align}
    \dot{\Sigma}_{\text{p}}(r) = f_{\rm ice}(T) \frac{\epsilon}{d(r)} \frac{\dot{M}_{\text{peb}}}{2 \pi r}
    \label{planetesimal_column_density}
\end{align}
with $d(r)$ as the radial \hubert{separation} of the pebble traps and $\epsilon$ as the efficiency parameter, that describes how much of the pebble flux is transformed into planetesimals \hubert{after drifting over a distance of $d$}. We choose a constant value of $\epsilon=0.05$ as a good value to form a sufficient number of planetesimals as found in \citet{Lenz_2019} for $d(r)=5.0$ pressure scale heights\hubert{, motivated by our findings in the detailed numerical simulations of zonal flows \citep{Dittrich2012}}. Generally we can change $\epsilon$ locally, if the formation of planetesimals might follow a different underlying mechanism, like, e.g., around the water iceline as described by \cite{Dr_kowska_2017} or \cite{schoonenberg2017planetesimal}. This flexibility allows us to study a broad range of planetesimal formation scenarios, using the same implementation. 
\hubert{Right now our two-poppy implementation has no proper treatment of the processes of evaporation and possible recondensation. The only effect of the existing iceline is incorporated into the parameter $f_{\rm ice}(T)$:
\begin{align}
    f_{\rm ice}(T)= \left\{\begin{array}{ll} 
        1 & \text{for } \, T <= 170 {\rm K} \\
        \frac{1}{3} & \text{for }\, T > 170 {\rm K}
        \end{array}\right..
        \label{f_m_parameter}
\end{align}
in effect to reduce the pebble flux inside the iceline to compensate for the evaporation of water ice. Therefore the ice line is visible in the distribution of planetesimals, even so it is not visible in the pebbles themselves (See Fig.\ \ref{fig:disk_evolution}).
}
We also use a fixed planetesimal size of 100$\,$km in diameter as in \citep{Lenz_2019}. As a consequence, there is a threshold of transformed mass to be reached to build at least one planetesimal. From this we can derive a critical pebble flux, that is necessary for $\Sigma_P$ to change. It is given as \citep{Lenz_2019}
\begin{align}
    \dot{M}_\text{cr} := \frac{m_p}{\epsilon \tau_t}
    \label{critical_pebble_flux}
\end{align}
where $\tau_t$ describes the average lifetime of a trap, which is given as 100 local orbits and $m_P$ the mass of a single planetesimal. For simplicity we assume spherical planetesimals with a uniform density of $\rho_s =  1.0 \text{g/cm}^3$. The mass that is transformed into planetesimals arises as a sink term in the advection diffusion equation (Eq. ~\ref{advection_diffusion_equation}). The new advection-diffusion equation is then given as
\begin{align}
    \frac{\partial \Sigma_s}{\partial t} + 
    \frac{1}{r} \frac{\partial}{\partial r} 
    \left[ 
    r \left(
    \Sigma_s \bar{u} - D_{g} \Sigma_g \frac{\partial}{\partial r}
    \left(
    \frac{\Sigma_s}{\Sigma_g}
    \right)
    \right)
    \right]
    = L
    \label{advection_diffusion_equation_with_L}
\end{align}
where the sink term $L$ is defined as 
\begin{align}
    L = (1-f_m(r)) \cdot L_0 + f_m(r) \cdot L_1
    \label{sinkterm}
\end{align}
with
\begin{align}
    L_0 = \frac{\epsilon}{d(r)} 
    \cdot v_\text{drift 0} \Sigma_0
    \cdot \theta \,(\dot{M}_\text{peb}   - \dot{M}_\text{cr}) \nonumber
    \\
    \times \, \theta \, (\text{St}_\text{0}   - \text{St}_\text{min})
    \cdot  \theta \, (\text{St}_\text{max} - \text{St}_\text{0})
    \label{sinkterm_0}
\end{align}
and
\begin{align}
    L_1 = \frac{\epsilon}{d(r)} 
    \cdot v_\text{drift 1} \Sigma_1
    \cdot \theta \, (\dot{M}_\text{peb}   - \dot{M}_\text{cr}) \nonumber
    \\
    \times \, \theta \, (\text{St}_\text{1}   - \text{St}_\text{min})
    \cdot  \theta \, (\text{St}_\text{max} - \text{St}_\text{1}) \vspace{1cm}.
    \label{sinkterm_1}
\end{align}
where $\theta \, (\cdot)$ is the heavyside function. This combines the above mentioned conditions for planetesimal formation. The surface density can only change while a critical mass is transformed ($\theta \, (\dot{M}_\text{peb}   - \dot{M}_\text{cr})$) and if the Stokes numbers of the particles are within $\text{St}_\text{min}$ and  $\text{St}_\text{max}$ ($\theta \, (\text{St}_\text{0}   - \text{St}_\text{min}) \cdot \theta \, (\text{St}_\text{mas} - \text{St}_\text{0})$).
\begin{figure}
    \centering
    \vspace{0.2cm}
    \begin{flushleft}
    \textbf{Solid evolution stages of our formation model: \\
    Origin model without planetesimal formation \\
    \citep{mordasini2018planetary} } 
    \end{flushleft}
    \centering
    \fbox{
    \includegraphics[width=0.47\textwidth]{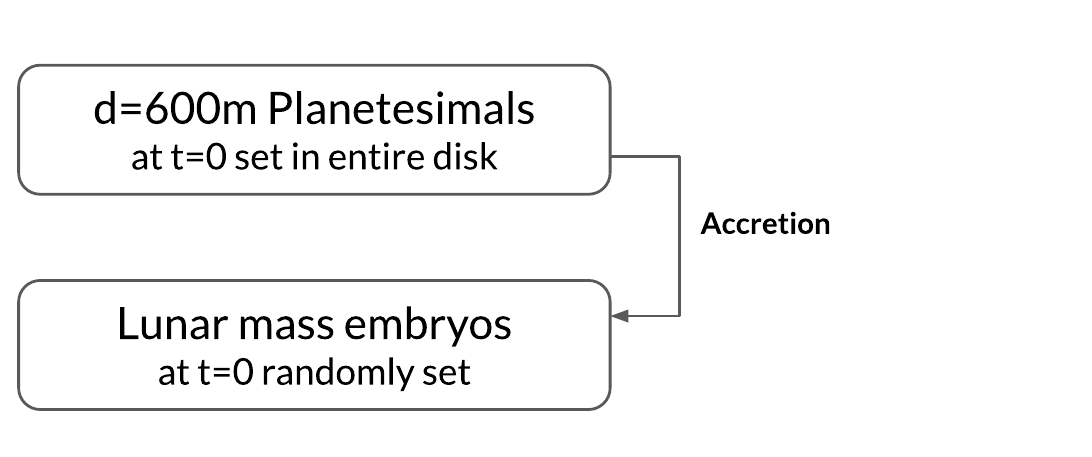}
    }
    \begin{flushleft}
    \textbf{Solid evolution as described in Sect. ~\ref{planetesimal_formation_model}} 
    \end{flushleft}
    \fbox{
    \includegraphics[width=0.47\textwidth]{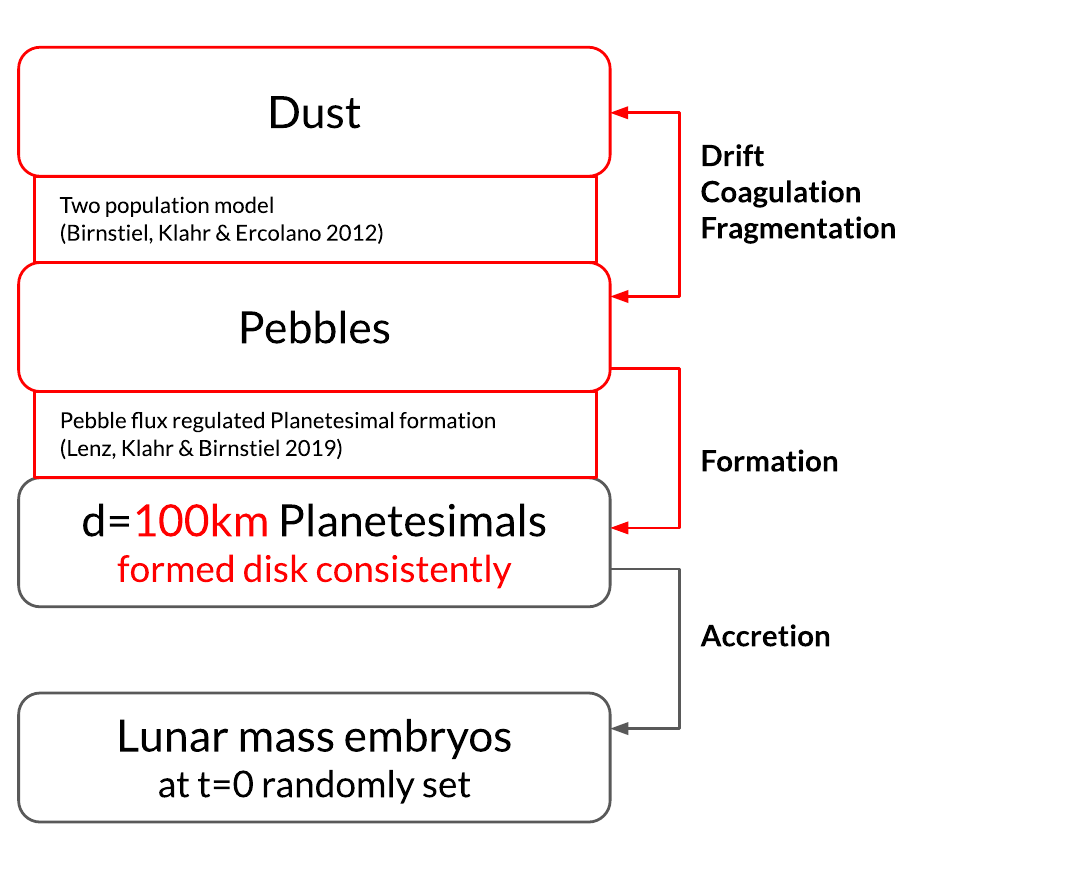}
    }
    \begin{flushleft}
    \textbf{Future possibilities}
    \end{flushleft}
    \fbox{
    \includegraphics[width=0.47\textwidth]{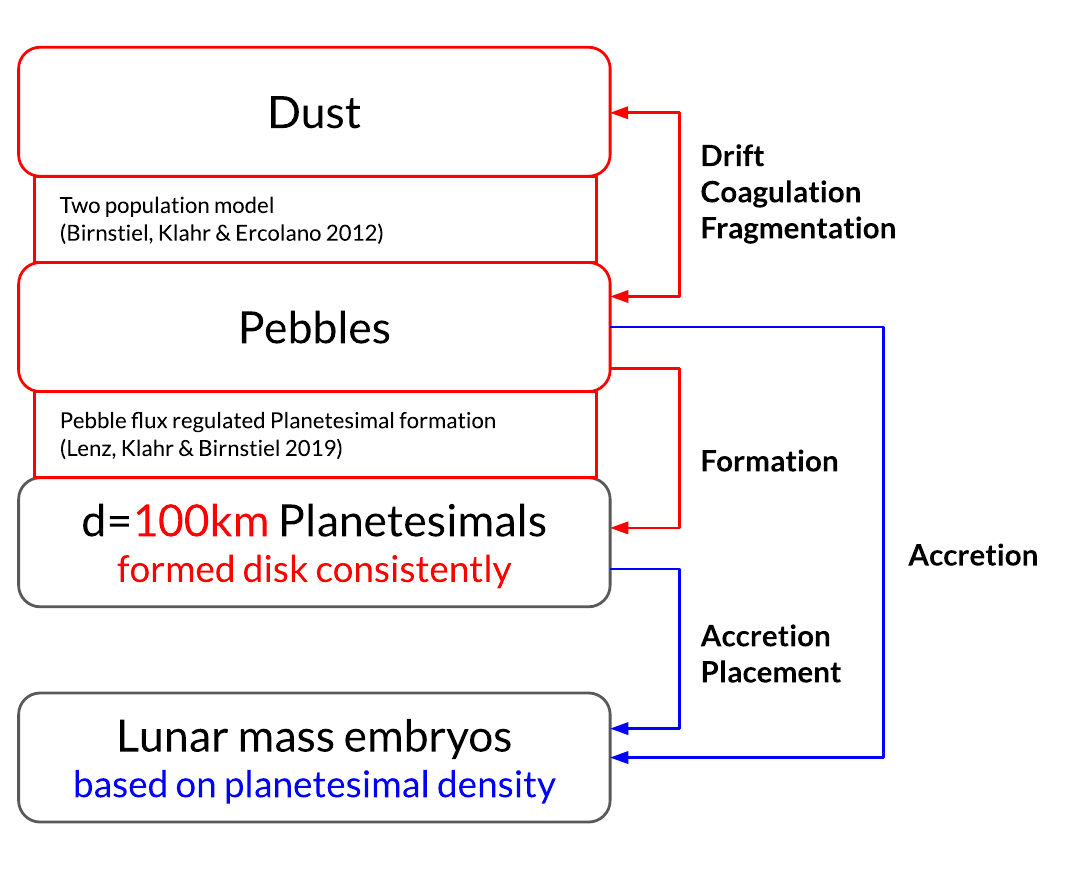}
    }
    \caption{Schematic display of the different formation model solid evolution development stages. The upper panel describes the previously published model from \cite{mordasini2018planetary}. The middle panel shows the currently improved version in this work, including the two population solid evolution for dust and pebbles, as well as the formation of planetesimals (see Sec. ~\ref{planetesimal_formation_model}). The lower panel gives an outlook on possible future development stages. The new modules and functions are highlighted in red, whereas future possibilities are highlighted in blue.}
    \label{fig:Bern_models}
\end{figure}

\section{The planet formation and evolution model}
\label{Bern_model_for_population_synthesis}

The most up to date version of our planet population synthesis model can be found in \citet{EmsenhuberPrepA}, which corresponds to an update of the model presented in \citet{mordasini2018planetary}. This model combines planet formation \citep{alibert2005models,alibert2013theoretical} and evolution \citep{mordasini2012combined}. Descriptions of the model can be found in \citet{benz2014planet}, \citet{mordasini2015global}, \citet{mordasini2018planetary}, and in upcoming work \citet{EmsenhuberPrepA}. We will provide here only an overview of the physical processes that are tracked in the model, while focusing on the solid components of the protoplanetary disk model in Sect.~\ref{Solid_component}.

The formation part of the model follows the core accretion scenario of planetary embryos in viscously-evolving circumstellar disks \citep{lust1952,lyndenbellpringle1974}. The macroscopic viscosity is given by the $\alpha$ parametrization \citep{shakura1973black}. Planetesimals are assumed to be in the oligarchic regime \citep{idamakino1993,thommes2003,chambers2006,Fortier2013}. The structure of the envelope is retrieved by solving the internal structure equations \citep{bodenheimerpollack1986}. During the initial phase, gas accretion is governed by the ability to radiate away the potential energy gained by the accretion of both solids and gas \citep{pollack1996formation,leechiang2015}. The efficiency of cooling increases with the planet's mass and once the gas accretion rate is limited by the supply of the gas disk, the planet contracts \citep{bodenheimer2000}.

Planets embedded in a gas disk will undergo migration \citep[e.g.,][]{baruteau2014}. The model uses the prescription of \citet{dittkrist2014}. For type~I migration it is based on the work by \citet{paardekooper2010} while for type~II planets move in equilibrium with the gas disk. The switch between the two follows the criterion of \citet{crida2006}.

The formation stage lasts for the entire life time of the protoplanetary disk, but at least 10$\,$Myr. Once this is passed, the model switches to the evolution stage \citep{mordasini2012combined} where the planets are followed until 10$\,$Gyr. This stage follows the thermodynamical evolution of the planets, with atmospheric escape \citep{lin2014} and tidal migration.

To perform population synthesis, we use a method similar to \citet{mordasini2009extrasolar}, with several adaptations. The distribution of disk gas masses and the relationship between the mass and the exponential cutoff radius follow \citet{Andrews_2010}. The inner radius is fixed to $0.03\,$au. The initial embryo mass is $0.0123\,M_\oplus$ and the location is random with a uniform distribution in the logarithm of the distance between 0.06 and 40$\,$au. Embryos are placed directly at the beginning of the simulations.

\subsection{The solid component}
\label{Solid_component}

A schematic overview of the different modules can be seen in Fig.~\ref{fig:Bern_models}. Previous generations of the model including the upcoming \citet{EmsenhuberPrepA} use an initial planetesimal surface density slope that was set either to be equal to the initial gas density slope \citep{mordasini2009extrasolar} or used a $\Sigma_P \propto r^{-1.5}$ mmsn-like distribution \citep{EmsenhuberPrepA}. For the first case, this gave a planetesimal surface density distribution of $\Sigma_P \propto r^{-0.9}$ up to an exponential cutoff radius, which depends on the given disk size. The total mass in planetesimals was chosen to be the metalicity (in the following dust to gas ratio $d_g$) of the host star times the total gas disk mass, modulo the effect of condensation fronts. The size of the planetesimals is chosen to be uniform and with a radius of $r_P=300 \,$m. Important to point out is that $\Sigma_P$ only evolved while being accreted or ejected by embryos. Planetesimal formation or drift were not included, which left us with a static distribution of planetesimals and a complete lack of a physical description of the early phases of planet formation. 

\hubert{With our newly implemented model for planetesimal formation we go beyond the standard implementation on \citet{EmsenhuberPrepA}. We now included two additional solid} quantities (dust $\&$ pebbles) that \hubert{are evolving along} with the gas evolution of the disk model. The initial mass in dust and pebbles is given as the metalicity of the host star times the gas disk mass. Their density slope is set to be equal to that of the gas disk, giving an initial solid density profile of $\Sigma_s \propto r^{-0.9}$. There are no initially placed planetesimals.
Planetesimals only form based on the evolution of dust and pebbles. This ensures that planetesimals form consistently with the disk evolution. Not only is the final distribution of planetesimals highly different than the static assumption of the previous disk model (see Sect.~\ref{Pebble flux regulated Planetesimal formation}) but also do planetesimals now form over time, which opens a completely new level of dynamical interaction with the disk. The size of planetesimals that we assume in the following simulations is given as $100\,$km in diameter.

\hubert{Thus the main differences between \citet{EmsenhuberPrepA} and this paper is the size of planetesimals ($r_P=300 \,$m vs. $r_P=50 \,$km) and the option for dynamic planetesimal formation, which is not yet implemented in \citet{EmsenhuberPrepA}. \citet{EmsenhuberPrepA} on the other side includes an N-body integration for multiple simultaneously evolving cores, an option we have not used in the present paper, as we wanted to focus on the effect of dynamical planetesimal formation.}

\section{Results}
\label{Results}
\begin{figure}
\centering
\includegraphics[width=0.49\textwidth]{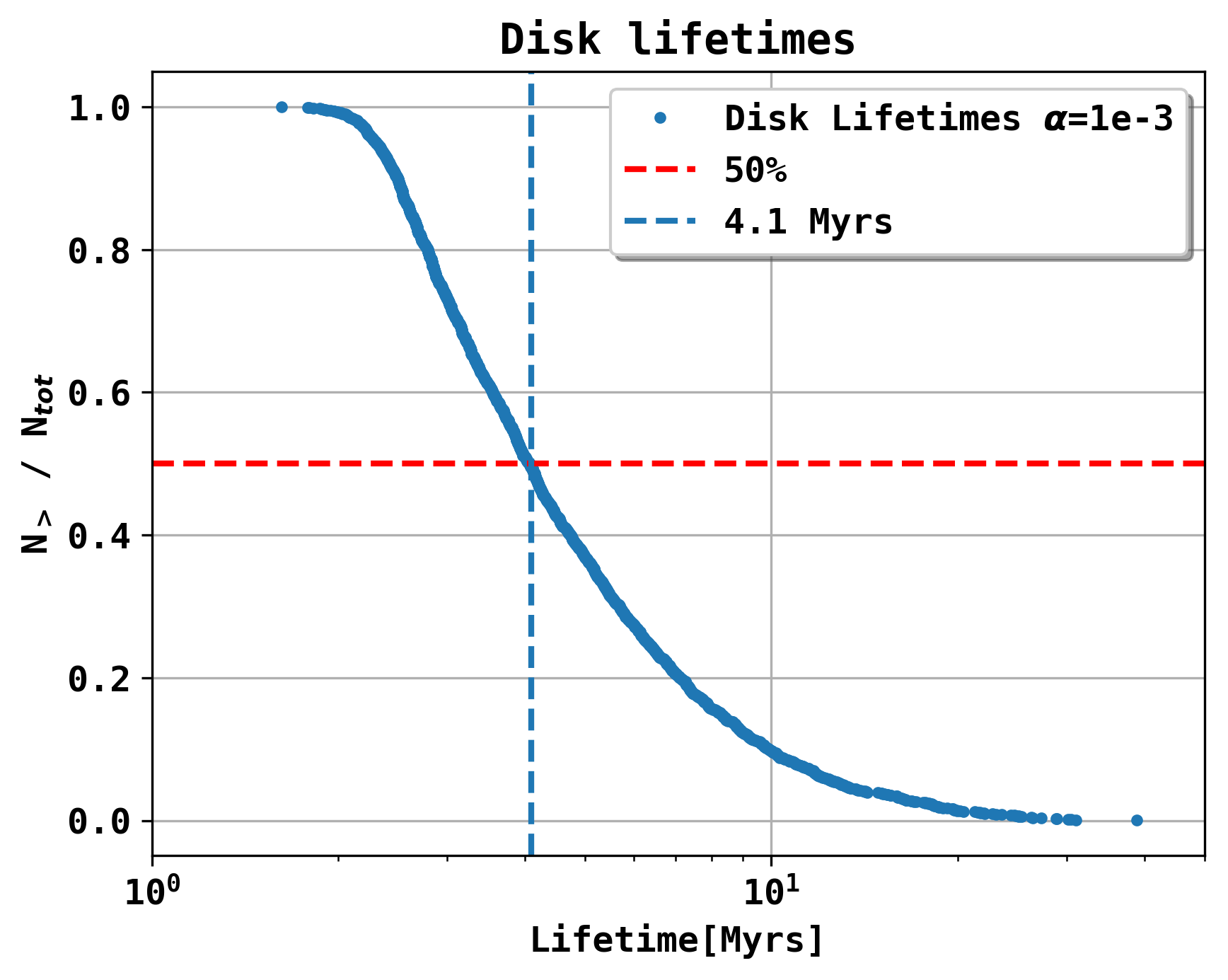}
\caption{Cumulative distribution of gas disk lifetimes for our synthetic population. We have used a constant value of $\alpha=10^{-3}$ in our runs. We find that 50$\%$ of the lifetimes are below 4.1 Myrs.}
\label{fig:disk_lifetime}
\end{figure}
\begin{figure*}
\centering
\begin{minipage}{0.49\textwidth}
\centering
\includegraphics[width=\textwidth]{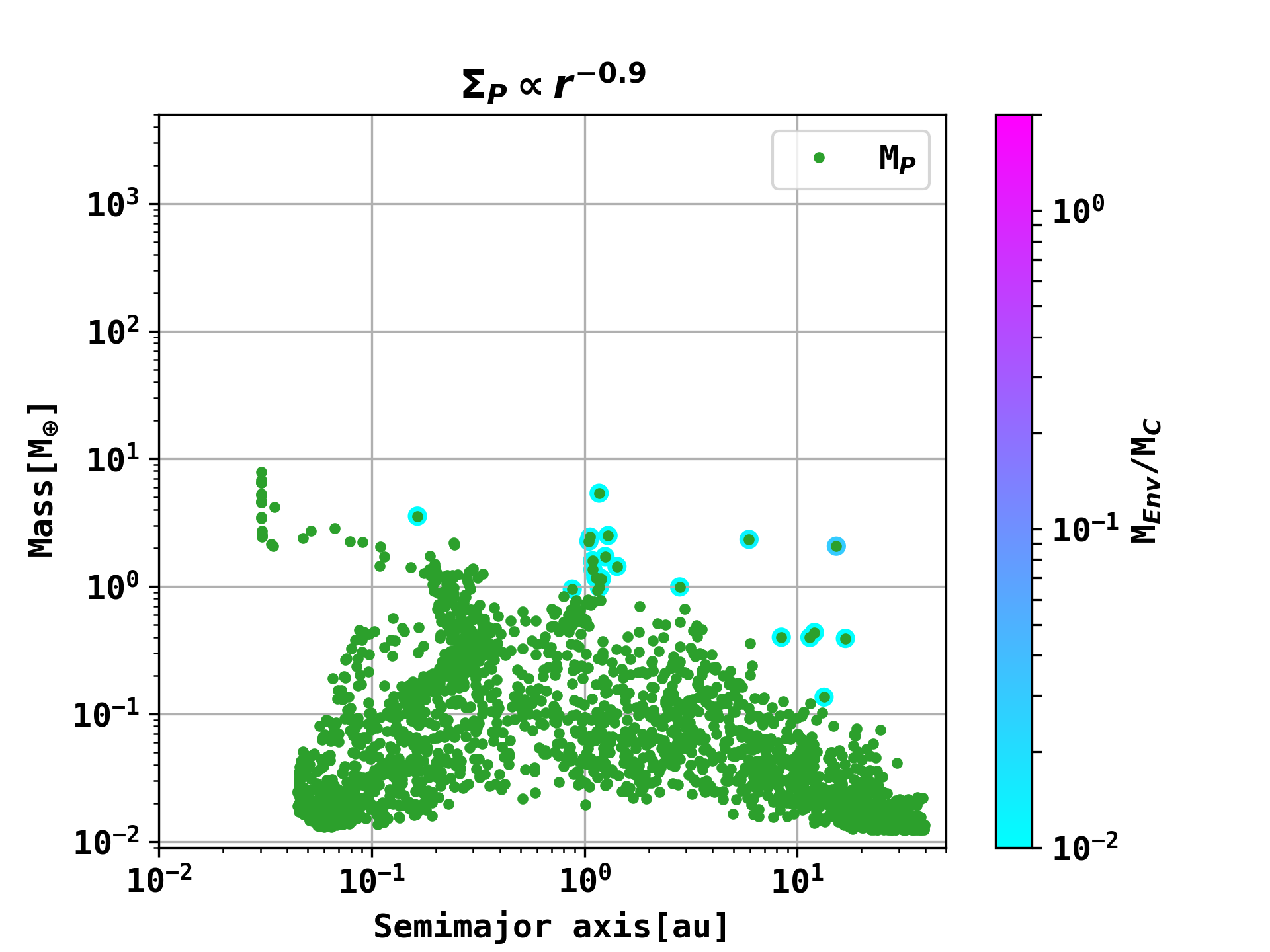}
\end{minipage}
\begin{minipage}{0.49\textwidth}
\centering
\includegraphics[width=\textwidth]{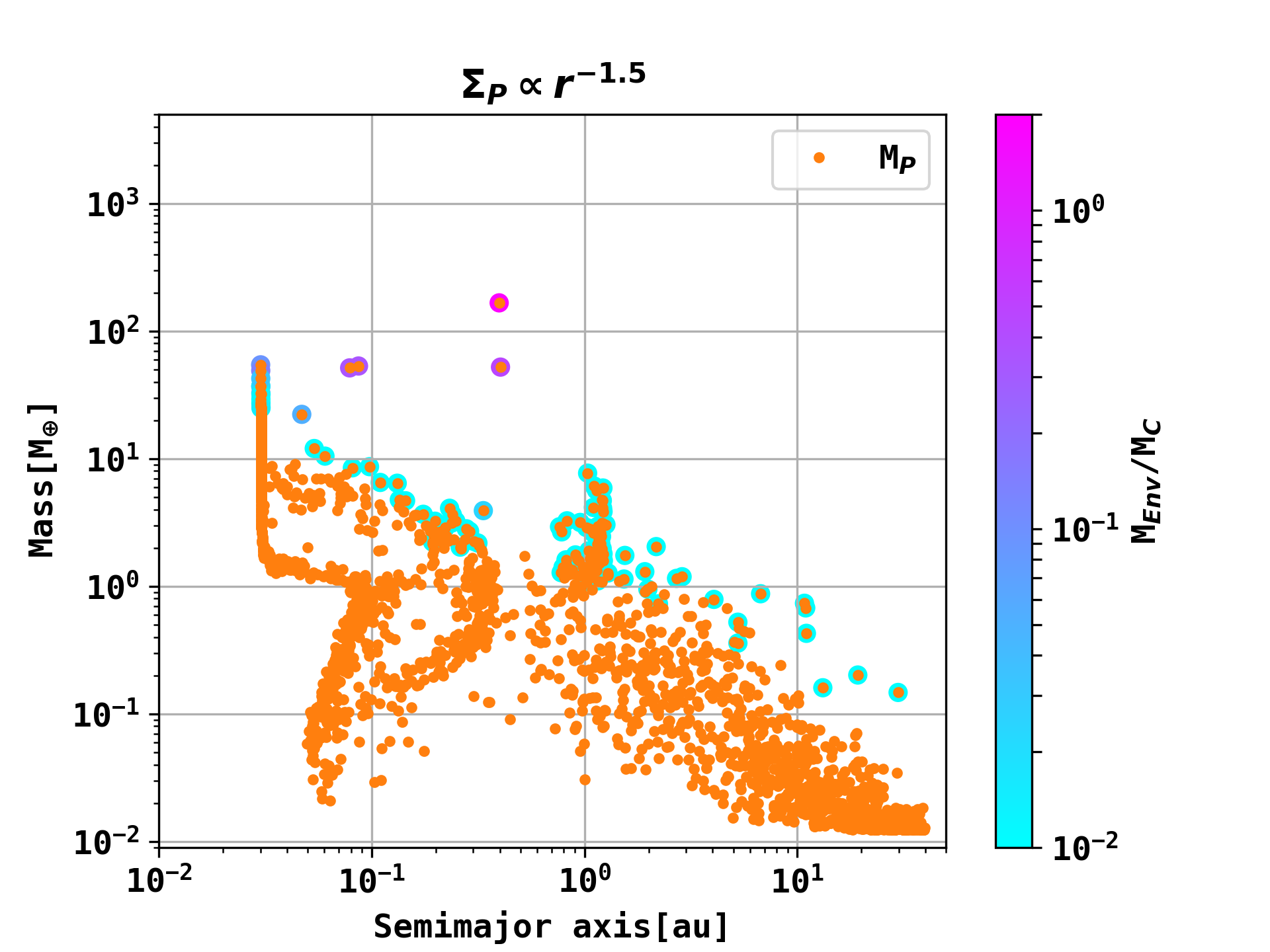}
\end{minipage}
\\
\centering
\begin{minipage}{0.49\textwidth}
\centering
\includegraphics[width=\textwidth]{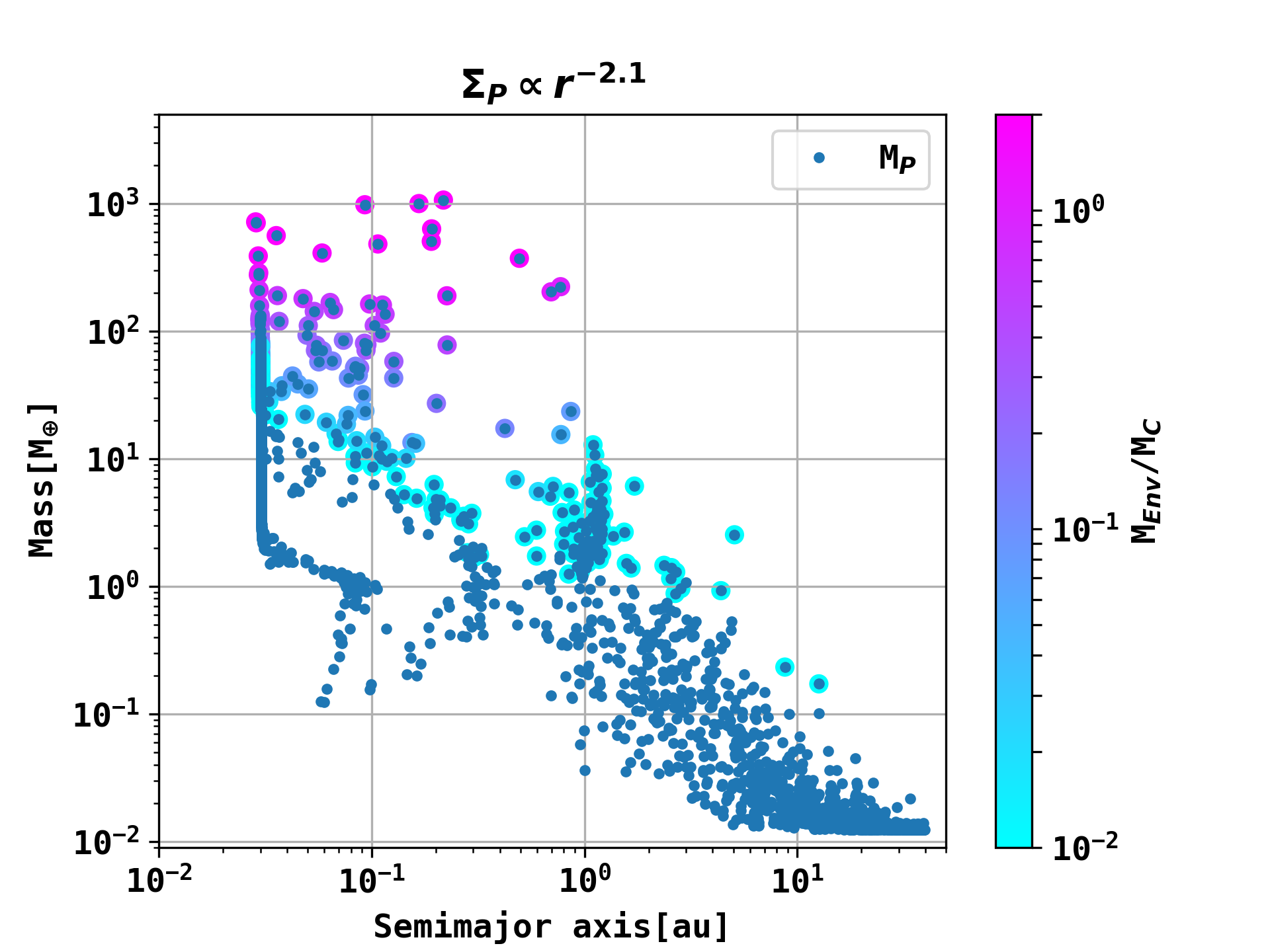}
\end{minipage}
\begin{minipage}{0.49\textwidth}
\centering
\includegraphics[width=\textwidth]{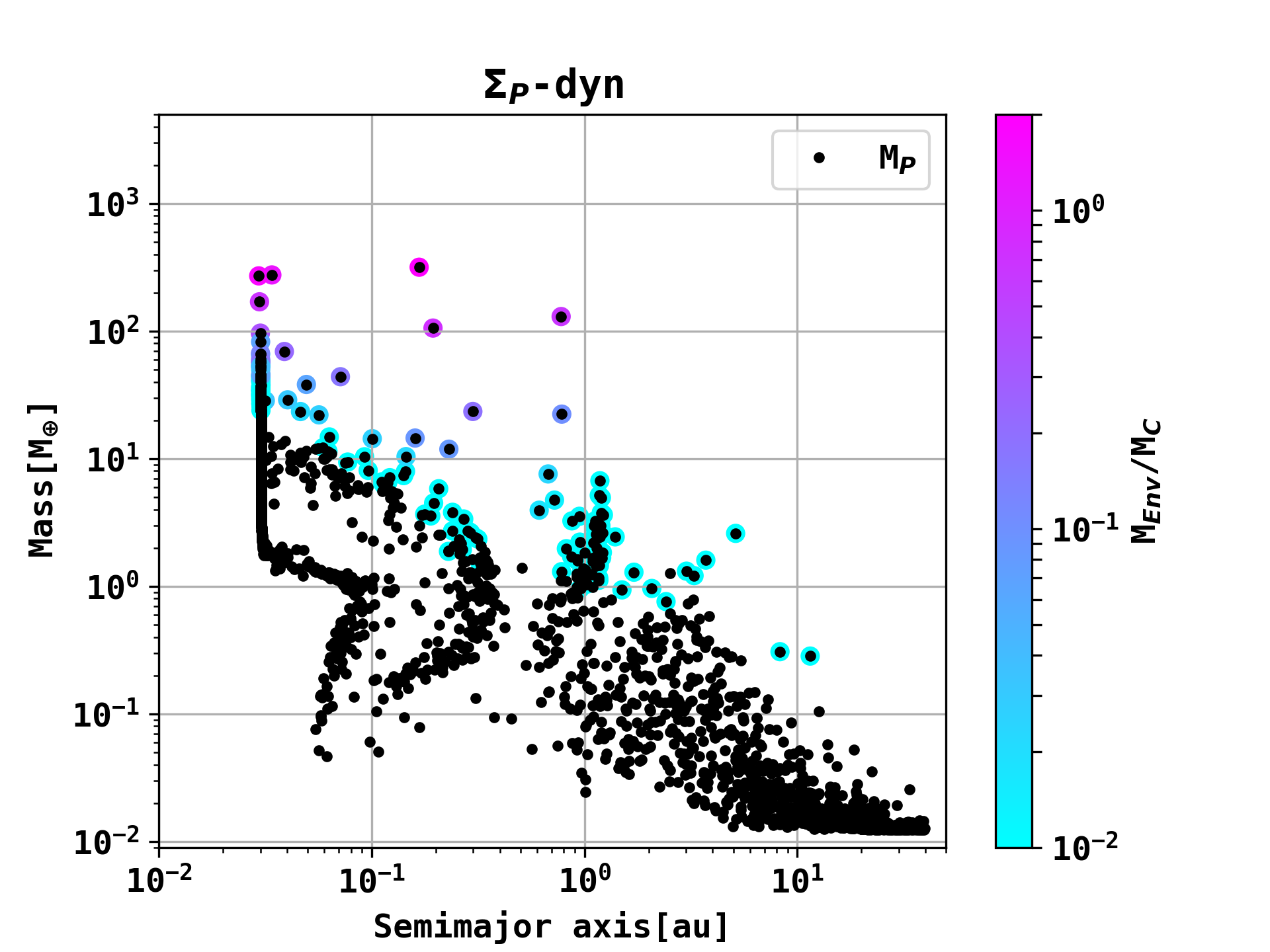}
\end{minipage}
\caption{
\small Mass versus semimajor axis of synthetic planet populations for different $\Sigma_P$ distributions after 100 million years. Each setup contains one single planetary embryo. The initially set distributions for $\Sigma_P$  are  $\Sigma_P = \Sigma_0 \cdot r^{-0.9}$ (initial gas density slope),  $\Sigma_P = \Sigma_0 \cdot r^{-1.5}$ (mmsn) and  $\Sigma_P = \Sigma_0 \cdot r^{-2.1}$ (Lenz et al 2019). The bottom right panel shows the population in which planetesimals form over time using the model described in Sec. ~\ref{planetesimal_formation_model}. The circles given around the datapoints show the mass fraction of envelope mass over core mass. The number of systems are 1999 ($\Sigma_P \propto r^{-0.9}$), 1990 ($\Sigma_P \propto r^{-1.5}$), 1961 ($\Sigma_P \propto r^{-2.1}$) and 1945 ($\Sigma_P$- dyn).
}
\label{Population_I}
\end{figure*}
\subsection{Disk evolution}
\label{Pebble flux regulated Planetesimal formation}
\begin{figure*}
\centering
\includegraphics[width=0.9\textwidth]{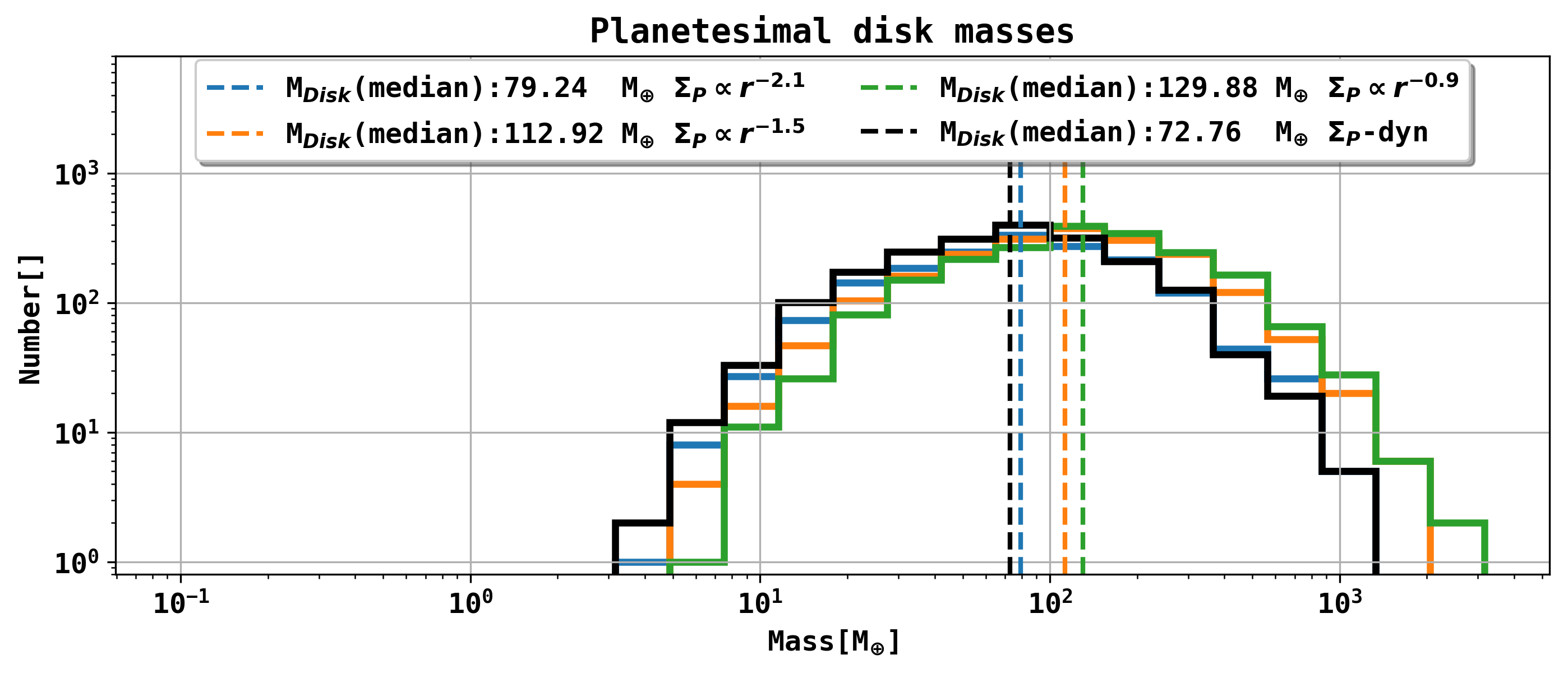}
\includegraphics[width=0.9\textwidth]{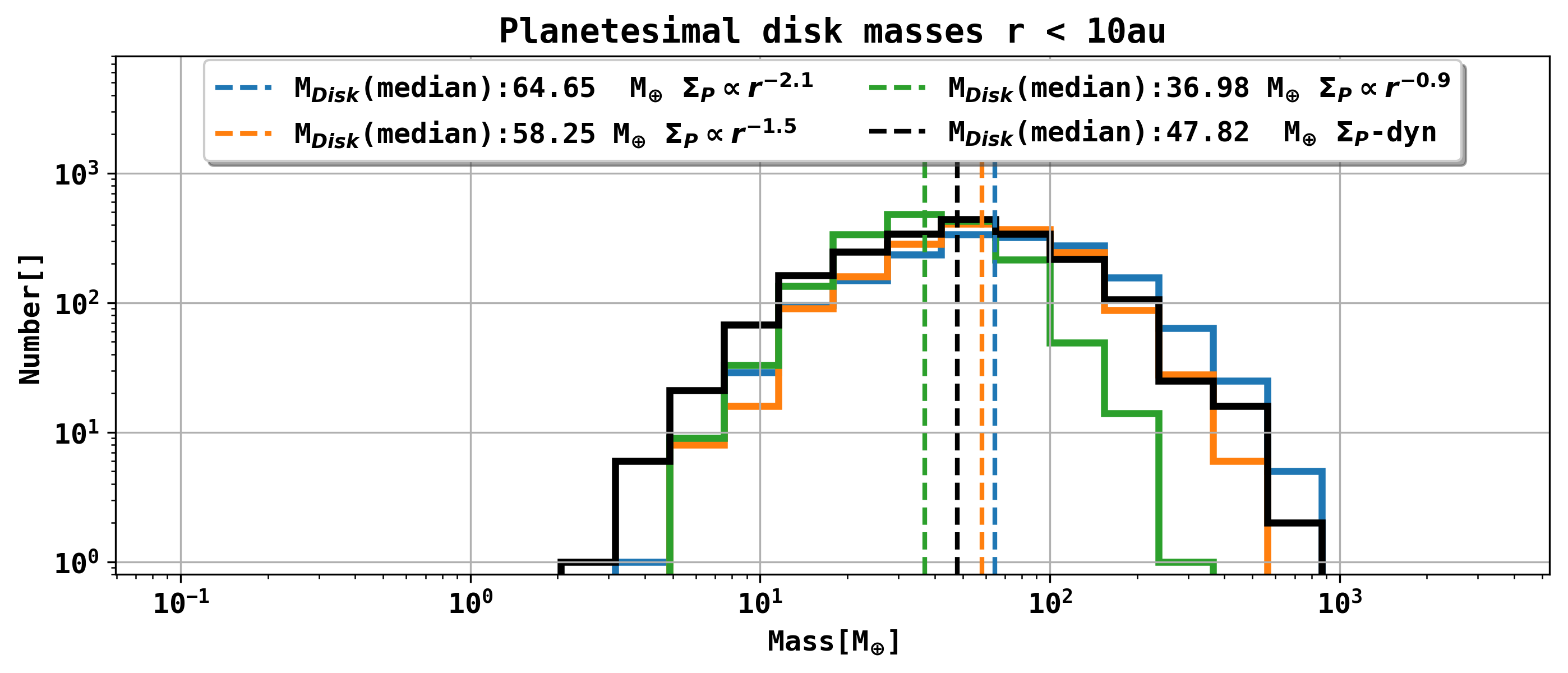}
\includegraphics[width=0.9\textwidth]{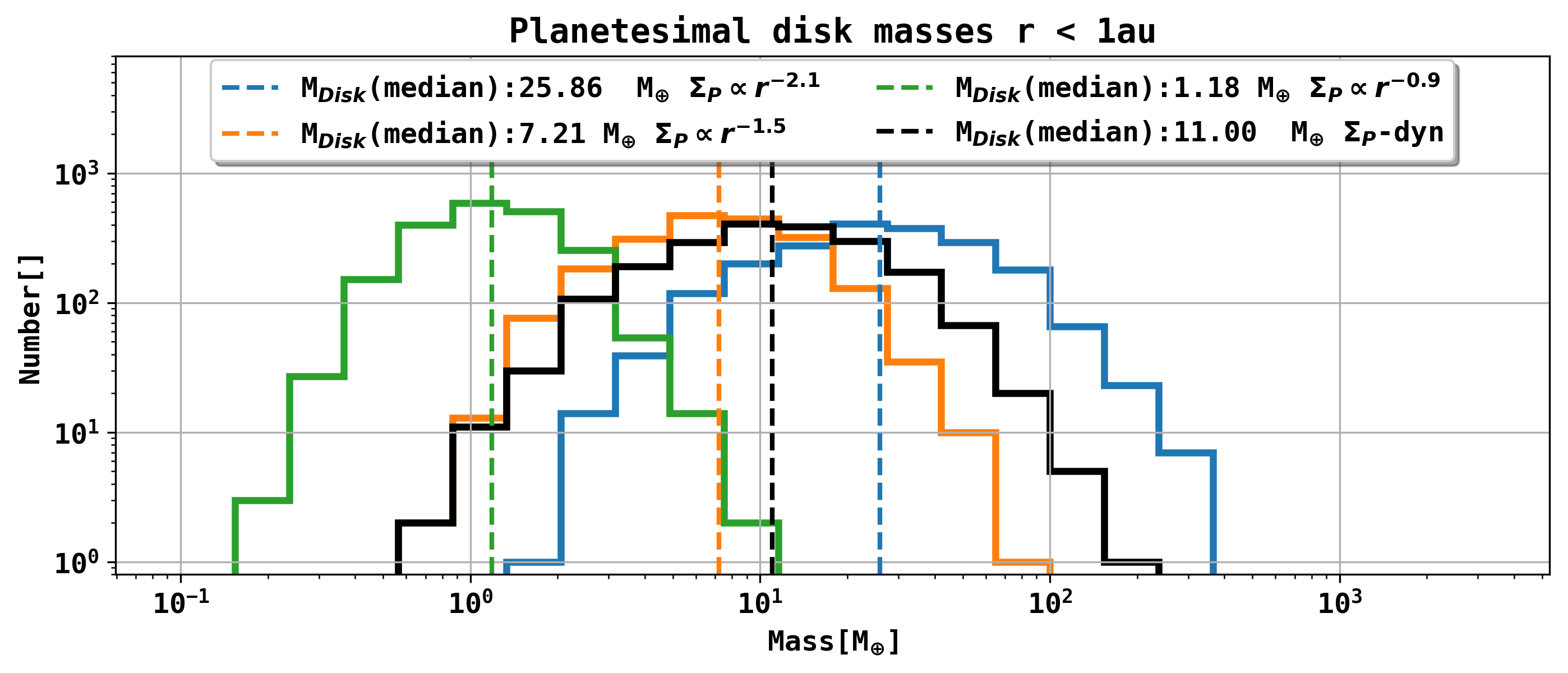}
\caption{Planetesimal disk masses within 1$\,$au, 10$\,$au and the complete disk for three different analytic density slopes and the dynamically formed planetesimal mass. The analytic masses are given at the start of the simulation while the dynamically formed disk masses are shown after one million years, after most planetesimals have already formed. The dynamic runs do not contain a planetary embryo, they only simulate the disk evolution. The disk parameters however are the same as in the population in Fig. ~\ref{Population_I}. We show the mass in planetesimals in the whole disk in the upper panel, the planetesimal mass within 10$\,$au in the middle panel, the planetesimal mass within 1$\,$au in the lower panel and the corresponding median masses for every setup.}
\label{fig:disk_masses}
\end{figure*}

Previous simulations with our model used an initial $\Sigma_P$ of $d_g \cdot \Sigma_g$ where $d_g$ is the dust to gas ratio. The slope in $\Sigma_P$ was therefore given as the slope of the initial gas surface density. \\
The density slope that arises from the pebble flux regulated model for planetesimal formation can have a slope as steep as $\Sigma_P \propto r^{-2.1}$, generally it depends on the individual evolution of the disk.
Due to the steeper slope, we find a remarkable increase of $\Sigma_P$ in the inner regions of a protoplanetary disk and a corresponding decrease further out. Another profound difference to the previous implementation of our model is the total mass in planetesimals.
The initial mass in dust in the planetesimal formation runs is equal to the initial mass in planetesimals with the analytically given planetesimal surface density, yet only a fraction of that is transformed into planetesimals. We will therefore always undershoot the total mass in planetesimals for our dynamically formed simulations, compared to the previous implementation. Choosing higher values for the planetesimal formation efficiency can result in a shallower density profile, similar to that of the initial gas distribution. The initial dust density is given as a fraction of the gas surface density. Considering $\epsilon/d > 1$, this would lead to \hubert{local} pebble-to-planetesimal \hubert{conversion} and the outer material could not drift into the inner regions of the disk, \hubert{which would have changed the density profile}. For a more detailed treatment of this behaviour we refer to \cite{Lenz_2019}. To find similar densities to the analytic $\Sigma_P \propto r^{-2.1}$ runs we would have to increase our disk masses to match the final mass in planetesimals. \\
The total disk masses for the different density distributions can be seen in Fig. ~\ref{fig:disk_masses}, as well as the masses within 10$\,$au and 1$\,$au. 

We find that the mean total disk masses are lower for the steeper density profiles by a factor of $M_{tot}^{-2.1} / M_{tot}^{-0.9} \approx 0.62 $ or $M_{tot }^{-1.5} / M_{tot}^{ -0.9} \approx 0.87 $. \hubert{This is to be expected as more material is inside the iceline, which is taken care of in these models.
Still the masses within 1$\,$au of the steeper models are by orders of magnitude higher ($M_{1au}^{ -2.1} / M_{1au }^{-0.9} \approx 21,58$ or $M_{1au }^{-1.5} / M_{1au }^{-0.9} \approx 6,01$).} $M_{tot}$ and $
M_{1au}$ refer to the median masses from Fig. ~\ref{fig:disk_masses}. The lowest total median mass ratios of planetesimals can be found in the dynamically formed simulation with \hubert{$M_{tot}^{\rm dyn} / M_{tot}^{ -0.9} = 0.504$, the mass ratio within 1$\,$au however is the second highest with $M_{1au}^{\rm dyn} / M_{1au}^{ -0.9} = 8.27$.} The smaller total masses for the steeper planetesimal surface density can be explained by the smaller amount of icy planetesimals in these setups.  Choosing a steeper density slope for the same mass as in the $\Sigma_P \propto r^{-0.9}$ shifts material (icy planetesimals and silicate planetesimals) from further out regions to the inner disk. This would evaporate the icy planetesimals within the iceline, leaving only the silicate planetesimals, therefore effectively loosing mass. The mass loss here is therefore only due to icy planetesimals within the iceline, whereas the amount of silicate planetesimals stays the same. Regardless of this mass loss, we find that the mass in the inner disk (r$\,$<$\,$1$\,$au) is signigicantly higher for the steeper density slopes. The lifetimes of the gas disks studied in our case can be seen in Fig. ~\ref{fig:disk_lifetime}
The global effect on planet formation of these changes in $\Sigma_P$ is presented in Sect. ~\ref{Population synthesis}.
\subsection{Synthetic populations}
\label{Population synthesis}
In the following we will present several synthetic populations that have been computed with different initial planetesimal surface density profiles and the dynamic planetesimal formation model from \cite{Lenz_2019}. It is important to mention that the growth of planetary embryos by the accretion of solids is only given by the accretion of planetesimals in these simulations. To ensure the right comparison of planetesimal accretion with different slopes of $\Sigma_P$, we neglect the accretion of pebbles for this part of our study. We also consider systems with one embryo each, because our focus lies on the changes to the previous implementation. Although populations with a much higher number of embryos are possible \hubert{in the new version of the model \citep{EmsenhuberPrepA}}, we chose to stay with \hubert{1 embryo per run for our study, as it makes no sense to mix our study with effects of multiple planets in that forthcoming paper. Therefore, }
we focus on the \hubert{general} distribution of masses and semimajor axes and the overall mass occurrences of planets.
\subsubsection{Mass semimajor axis distributions}
Figure ~\ref{Population_I} shows the mass and semi major axis distribution of four synthetic populations around a solar type star using 1 lunar mass (0.0123$\,$M$_{\oplus}$) planetary embryo for each system. We simulate a total number of 1999 systems for the $\Sigma_P \propto r^{-0.9}$ distribution, 1990 for $\Sigma_P \propto r^{-1.5}$, 1961 for $\Sigma_P \propto r^{-2.1}$ and 1945 for the dynamic planetesimal formation run. The initial conditions of the four populations are the same, except for the initial $\Sigma_P$ and the formation of planetesimals respectively. \\
The upper left green panel refers to an initial $\Sigma_P$ of $\Sigma_P \propto r^{-0.9}$, the upper right orange to $\Sigma_P \propto r^{-1.5}$ and the lower left blue to $\Sigma_P \propto r^{-2.1}$. The lower right panel in black refers to the final planets that formed using the pebble flux regulated model of planetesimal formation. 
We find a large number of planets that exceed a mass of ten earth masses (necessary for runaway gas accretion, see \citealp{pollack1996formation}) and sometimes even reach several hundreds of earth masses when the slope of $\Sigma_P$ is given with a slope of $r^{-2.1}$. The simulation in which the slope is given with the $r^{-0.9}$ does not even produce one single planet with a mass higher than that of ten earth masses. Overall this plot shows an immense increase in planetary masses for steeper planetesimal density profiles. It is important to mention here that the heavy gas giant planets all end up within 1$\,$au, which is due to the high masses in planetesimals in the inner disk and planetary migration. In our synthetic runs we do not see gas giants further out like e.g. beyond the water iceline as it can be observed in the population of exoplanets \citep{winn2015occurrence}\hubert{, which will probably change once we will incorporate recondensation of water vapor, effectively boosting the pebble flux regulated birth of planetesimals.}
\begin{figure}
    \centering
    \begin{minipage}{0.49\textwidth}
    \centering
    \includegraphics[width=\textwidth]{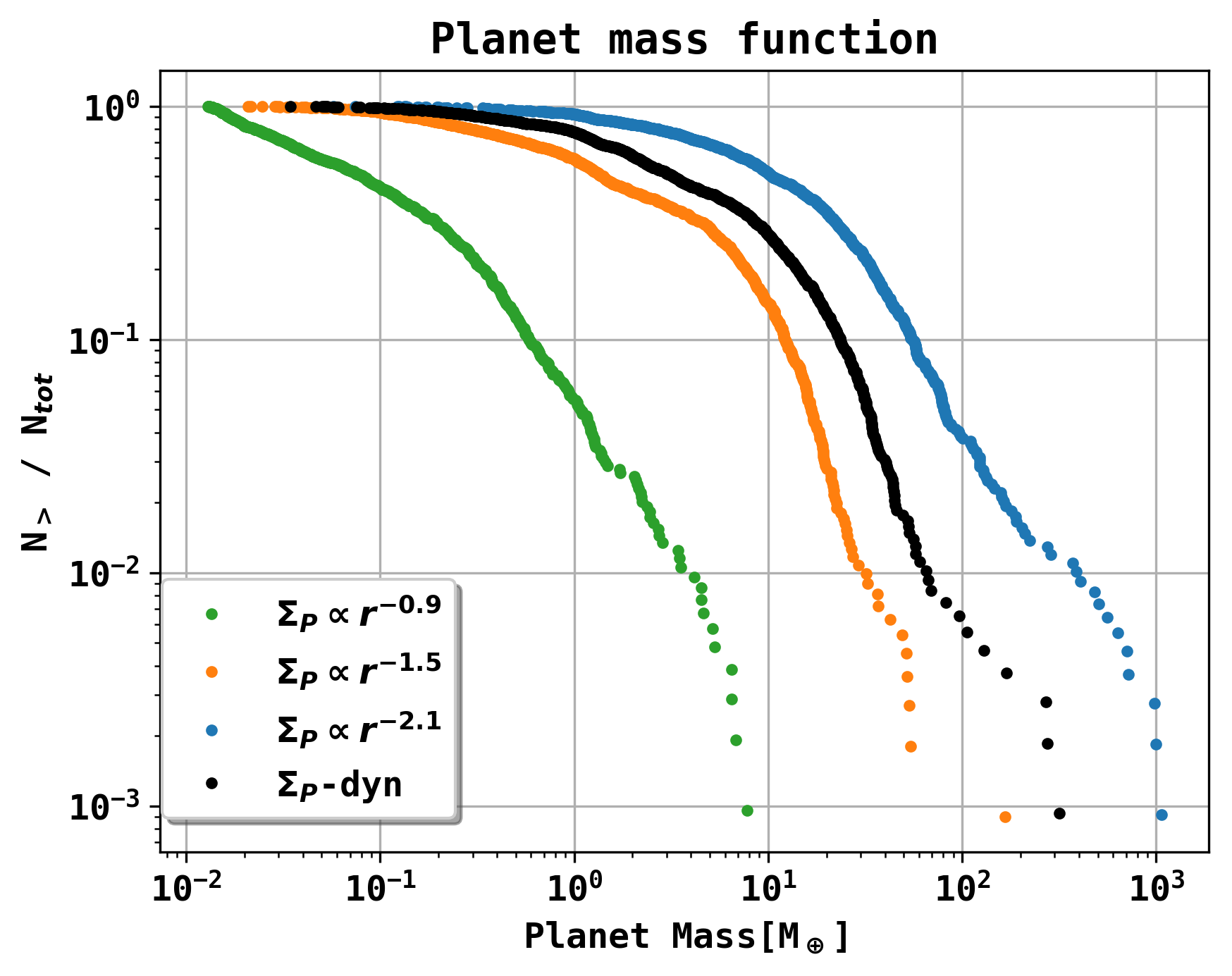}
    \end{minipage}
    \caption{Cumulative function of planetary masses within 1$\,$au for the four different synthetic populations from Fig. ~\ref{Population_I}. The y-axis shows how many planets for each density profile are above the current planetary mass, normalised by the total number for planets in each population. The planets that have formed in the $\Sigma_P \propto r^{-0.9}$ run are shown in green, the $\Sigma_P \propto r^{-1.5}$ population is shown in orange and the $\Sigma_P \propto r^{-2.1}$ population is shown in blue. The dynamic planetesimal formation population is shown in black.}
    \label{fig:cumulative_mass}
\end{figure}
\subsubsection{Mass occurences}
\begin{figure*}
\centering
\includegraphics[width=0.9\textwidth]{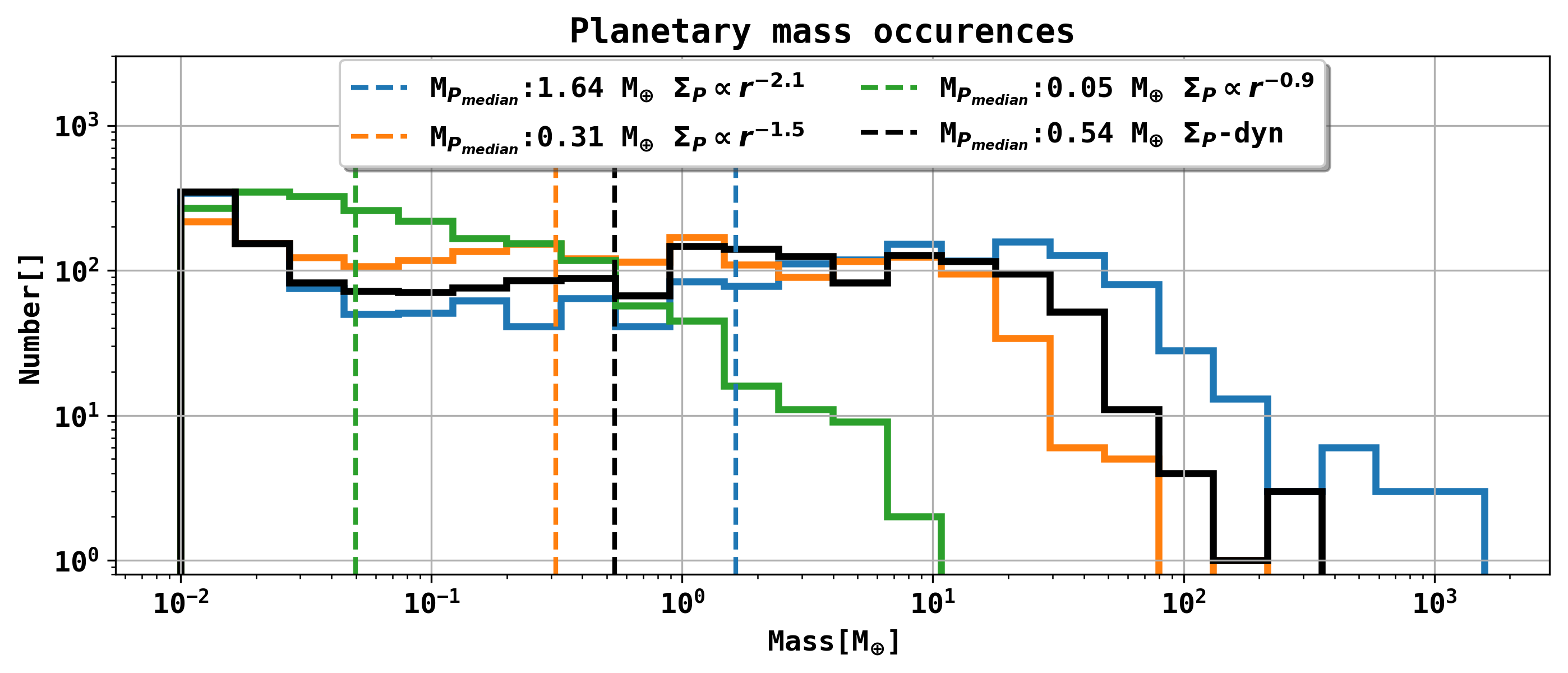}
\includegraphics[width=0.9\textwidth]{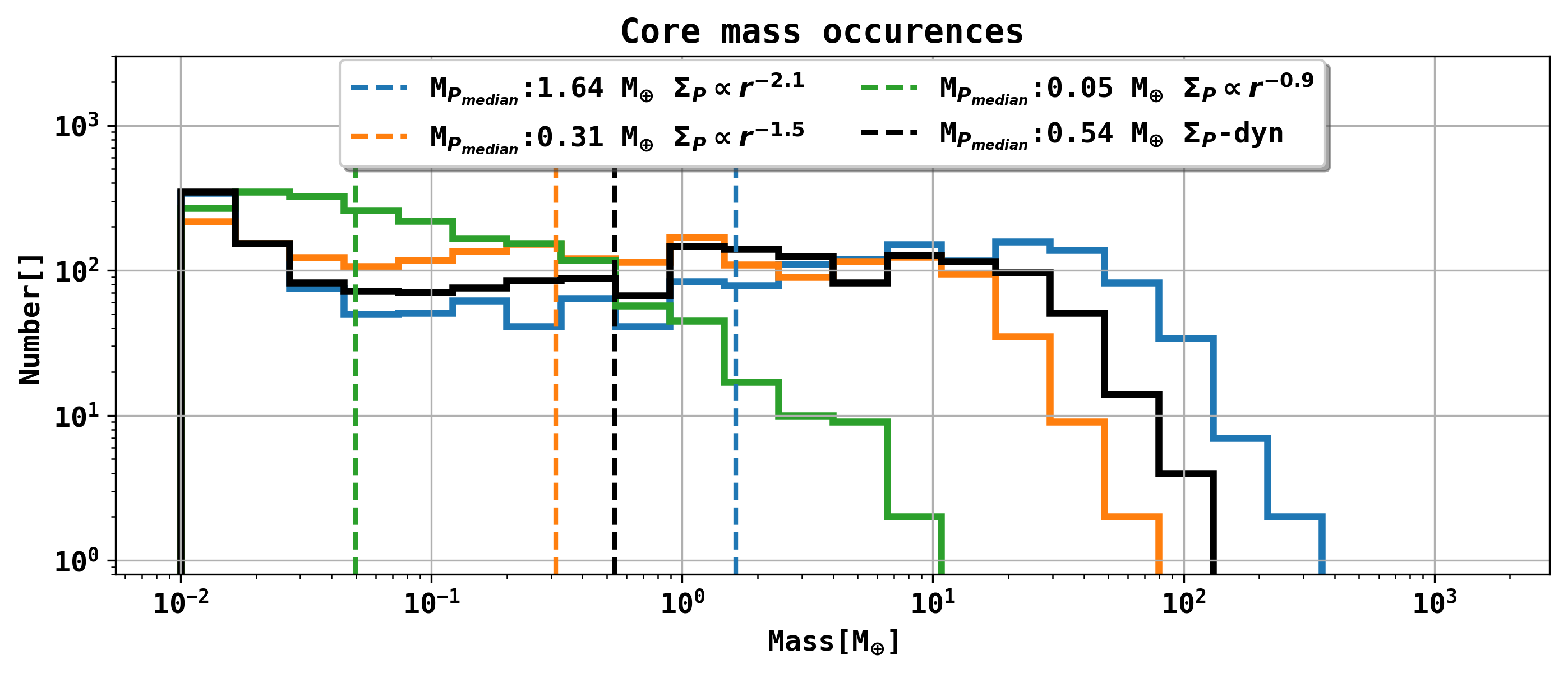}
\caption{Planetary and core mass occurences of the four different populations. Every planet in each systems starts with a core mass of  0.0123$\,$M$_\oplus$. and no envelope. The quantities that arise from the three analytical planetesimal surface density profiles are shown in in blue ($\Sigma_P \propto r^{-2.1}$), orange ($\Sigma_P \propto r^{-1.5}$) and green ($\Sigma_P \propto r^{-0.9}$), whereas the properties of the planetesimal formation population are shown in black. The dashed lines in the plots show the median planet and median core masses. The histograms show clear shifts towards the higher mass ranges for steeper planetesimal surface densities and for the dynamically formed planetesimals, compared to the $\Sigma_P \propto r^{-0.9}$ or even the $\Sigma_P \propto r^{-1.5}$ distribution.}
\label{fig:Total_mass_Histogramms}
\end{figure*}

\begin{figure*}
\centering
\includegraphics[width=0.9\textwidth]{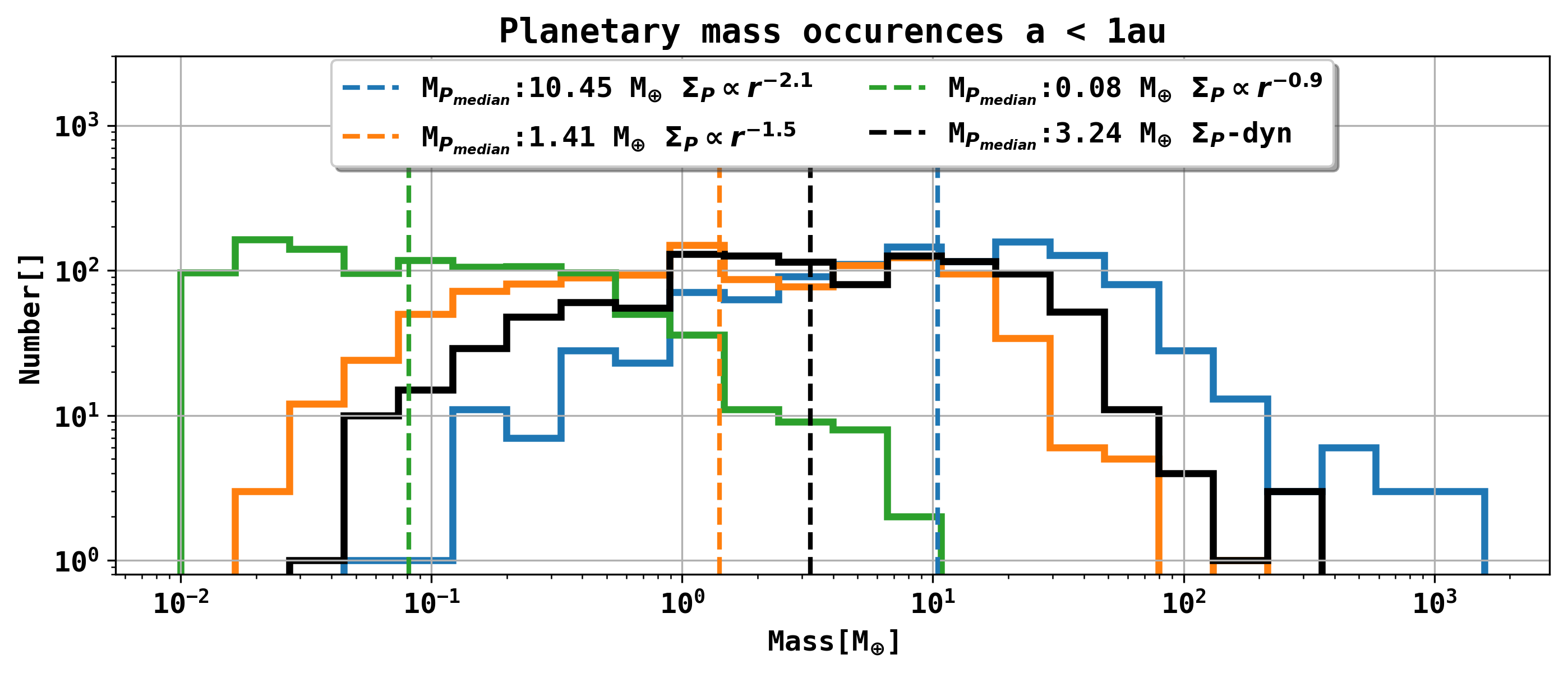}
\includegraphics[width=0.9\textwidth]{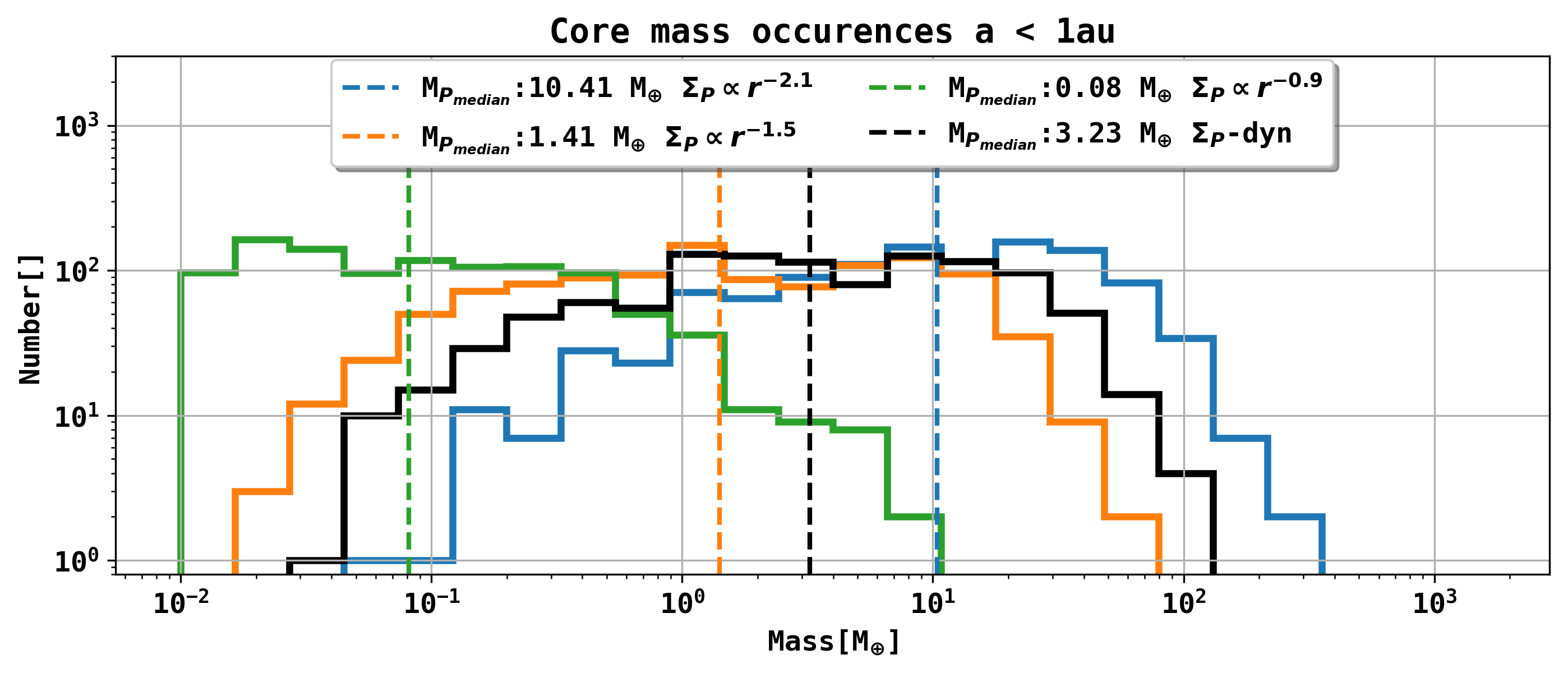}
\caption{Planetary mass and core mass occurrences of the four different populations within 1$\,$au. From Fig. ~\ref{Population_I} we see that the heavier planets are located in the inner disk since these are the regions with the highest planetesimal surface density. The dashed lines in the plots show the median planet and median core masses within 1$\,$au. We choose to focus on the planetary masses within 1$\,$au to neglect the "failed" planetary cores that are randomly placed at far out regions of the disk and do not grow substantially from the initial lunar mass.}
\label{fig:Total_mass_Histogramms_1au}
\end{figure*}

For a more quantitative analysis we study the mass occurrences for the different planetesimal density slopes. Here we focus on the planetary mass and the core mass. Fig. ~\ref{fig:Total_mass_Histogramms} and Fig. ~\ref{fig:Total_mass_Histogramms_1au} show histograms with the occurrences of the different masses for the various populations from Fig. ~\ref{Population_I}. As Fig. ~\ref{Population_I} shows, most of the \hubert{high} masses are found in the inner parts of the protoplanetary disk, whereas the outer placed embryos fail to grow. We therefore also focus our study on the inner region within 1$\,$au. Fig. ~\ref{fig:Total_mass_Histogramms} takes the complete population into account whereas Fig. ~\ref{fig:Total_mass_Histogramms_1au} only contains planets with a semimajor axis below 1$\,$au. We also give the median  masses for the planets and their cores. A cumulative function of the planetary masses is shown in Fig. ~\ref{fig:cumulative_mass}. We find that the number of planets above 10$\,$M$_\oplus$ is given as 0 ($\Sigma_P \propto r^{-0.9}$), 159 ($\Sigma_P \propto r^{-1.5}$), 565 ($\Sigma_P \propto r^{-2.1}$) and 301 ($\Sigma_P$- dyn). The number of planets above 20$\,$M$_\oplus$ is given as 31 ($\Sigma_P \propto r^{-1.5}$), 383 ($\Sigma_P \propto r^{-2.1}$) and 138 ($\Sigma_P$- dyn).
%%%%%%%%%% Starting location
\begin{figure}
    \centering
    \includegraphics[width=0.49\textwidth]{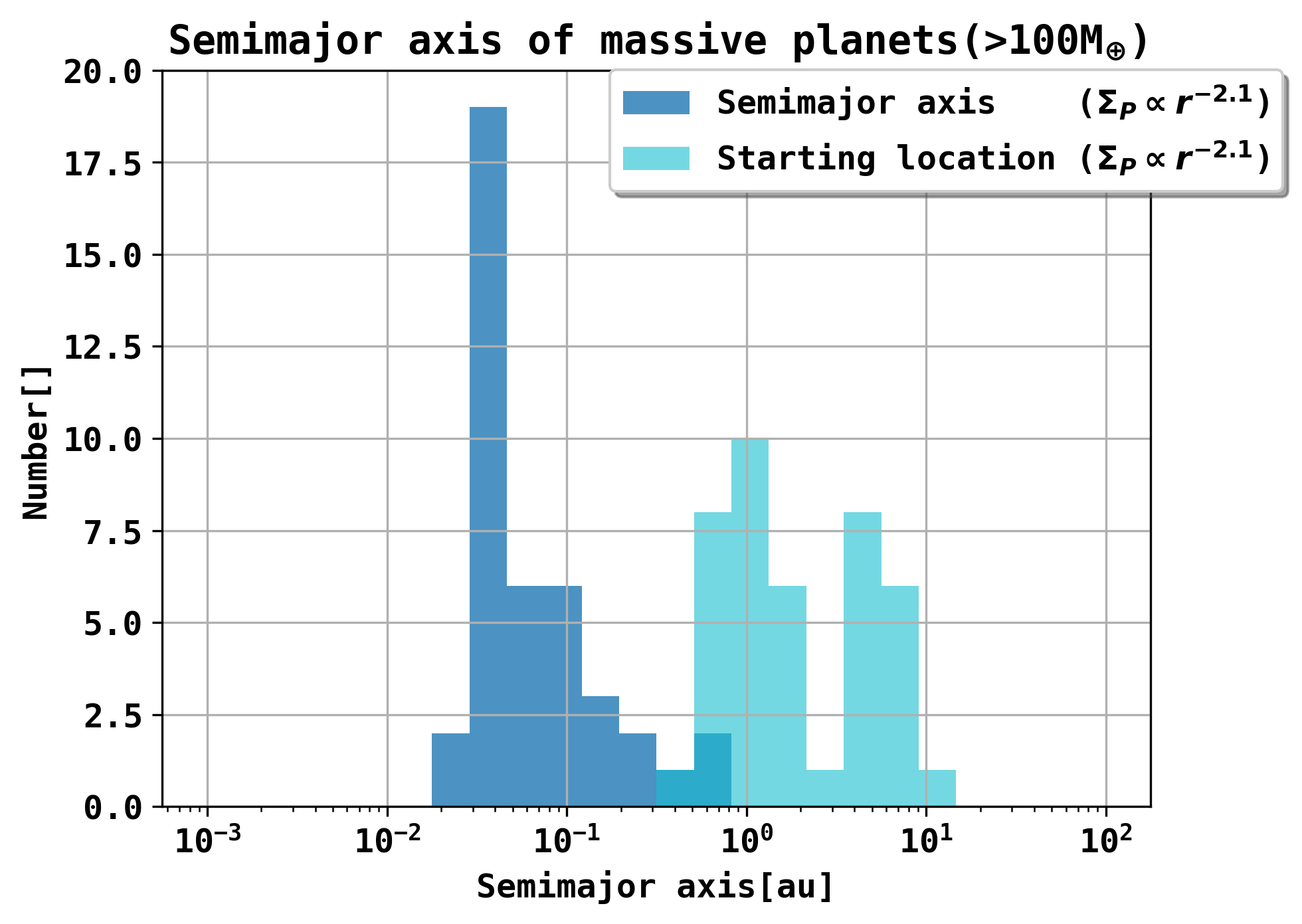}
    \\
    \centering
    \includegraphics[width=0.49\textwidth]{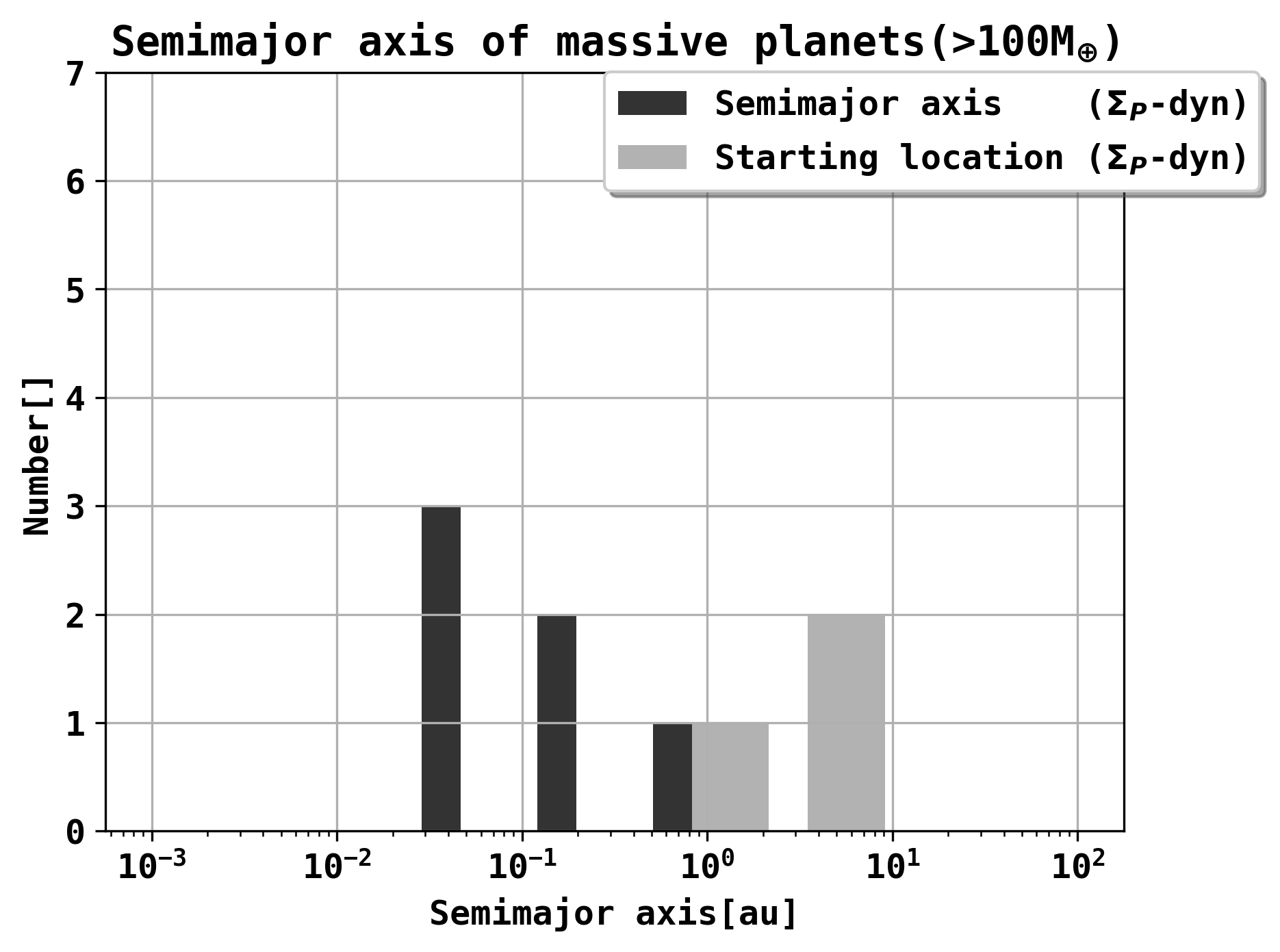}
    \caption{Semimajor axis distribution and starting location of planets that have grown to gas giant masses ($M_P > 100$M$_\oplus$) in the $\Sigma_P \propto r^{-2.1}$ and the planetesimal formation runs. We find that most \hubert{massive} planets end up at the inner edge of the disk. The total number of planets that have reached over 100M$_\oplus$ in the $\Sigma_P \propto r^{-2.1}$ runs is given as 41 out of 1961. This however is heavily biased by the placement of the planetary seeds which, also occurs at far out regions with low planetesimal surface densities. }
    \label{fig:Starting_location}
\end{figure}
\subsection{Influence of the starting location}
In Fig. ~\ref{fig:Starting_location} we see the semimajor axis distribution and the initial starting location distribution of high mass planets in the $\Sigma_P \propto r^{-2.1}$ run and the dynamic formation model. We find that most heavy planets end up at the inner edge of the disk due to migration. There is no in situ giant formation but rather a preferential zone in which planetary embryos need to be placed, in order to grow to giant planets. This preferential area appears to be around 1$\,$au and from around 4$\,$au to 10$\,$au for the $\Sigma_P \propto r^{-2.1}$ run and mostly from around 4$\,$au to 10$\,$au for the dynamical formation model. Embryos that \hubert{are placed} at a distance from 2$\,$au to 4$\,$au appear to have a lower probability to become gas giant planets in both cases, \hubert{but as the probability of their formation at this location is also low, due to the local deficiency in planetesimals, they should not have been placed there in the first place. Now that we have a distribution of planetesimals, we can use this information to model in the future also the generation of embryos in a consistent fashion. Ultimately the above mentioned effect of recondensation beyond the iceline can further change this picture.}
\subsection{Gas giant growth}
\begin{figure*}
\centering
\begin{minipage}{0.49\textwidth}
\centering
\includegraphics[width=\textwidth]{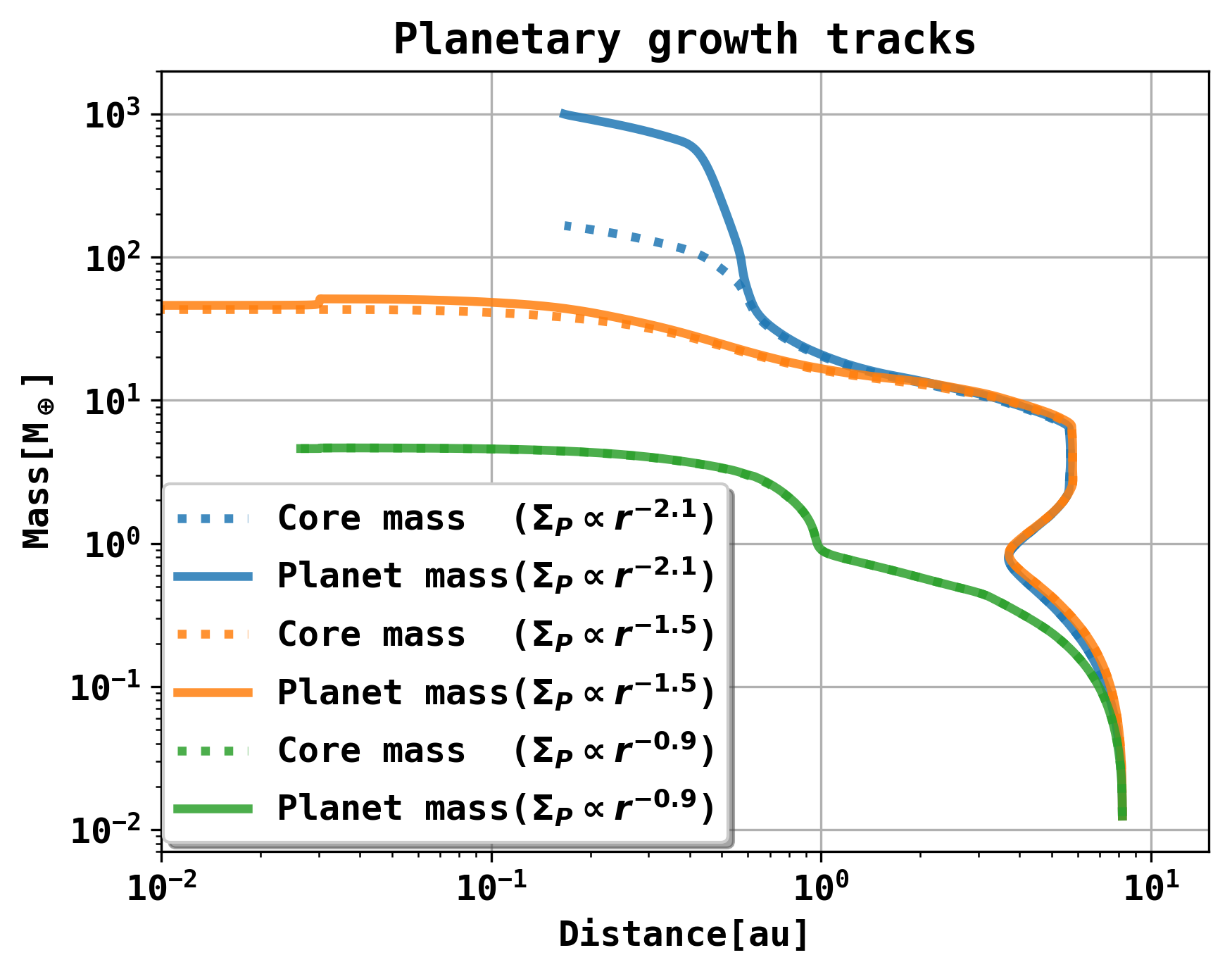}
\end{minipage}
\begin{minipage}{0.49\textwidth}
\centering
\includegraphics[width=\textwidth]{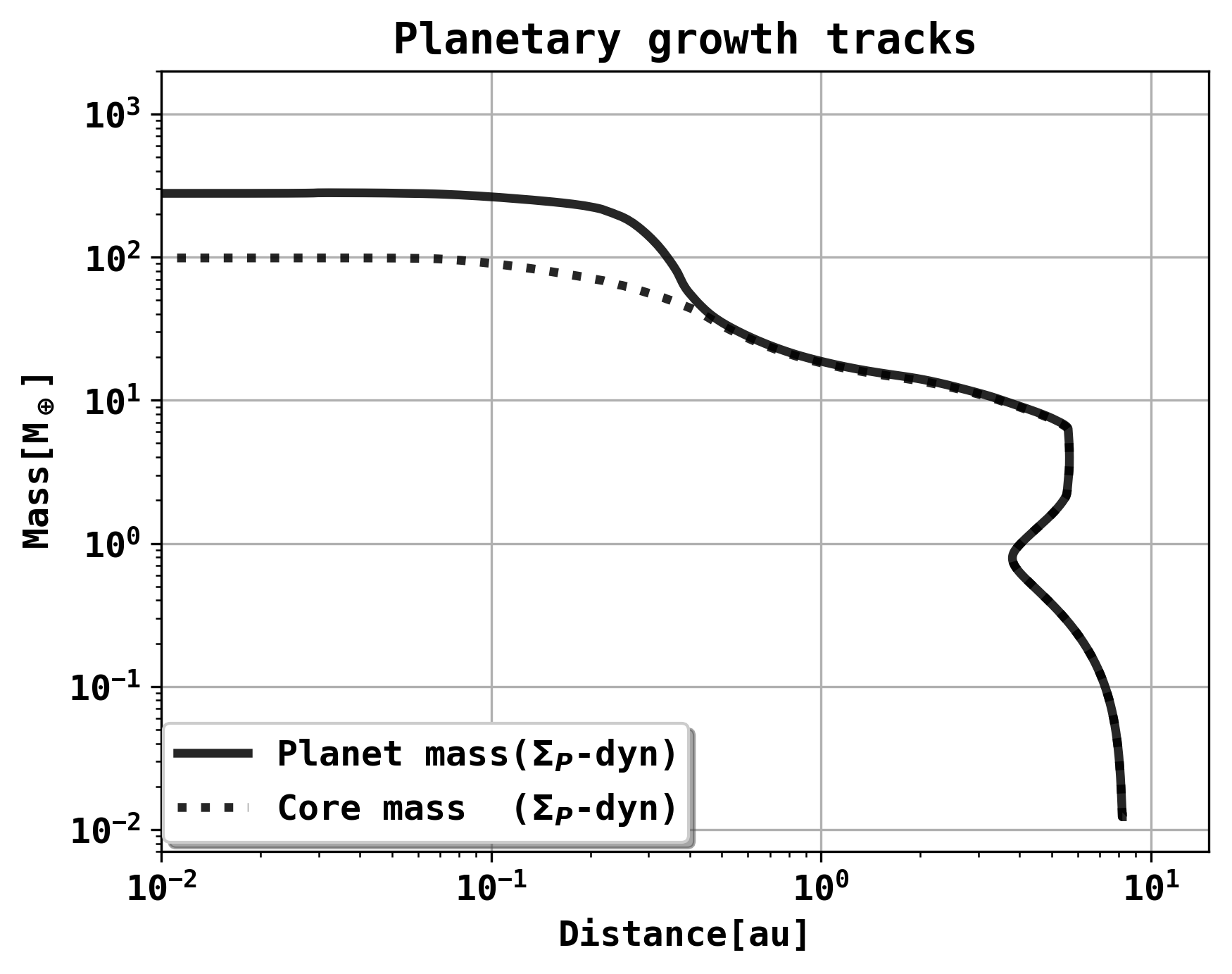}
\end{minipage}
\\
\centering
\begin{minipage}{0.49\textwidth}
\centering
\includegraphics[width=\textwidth]{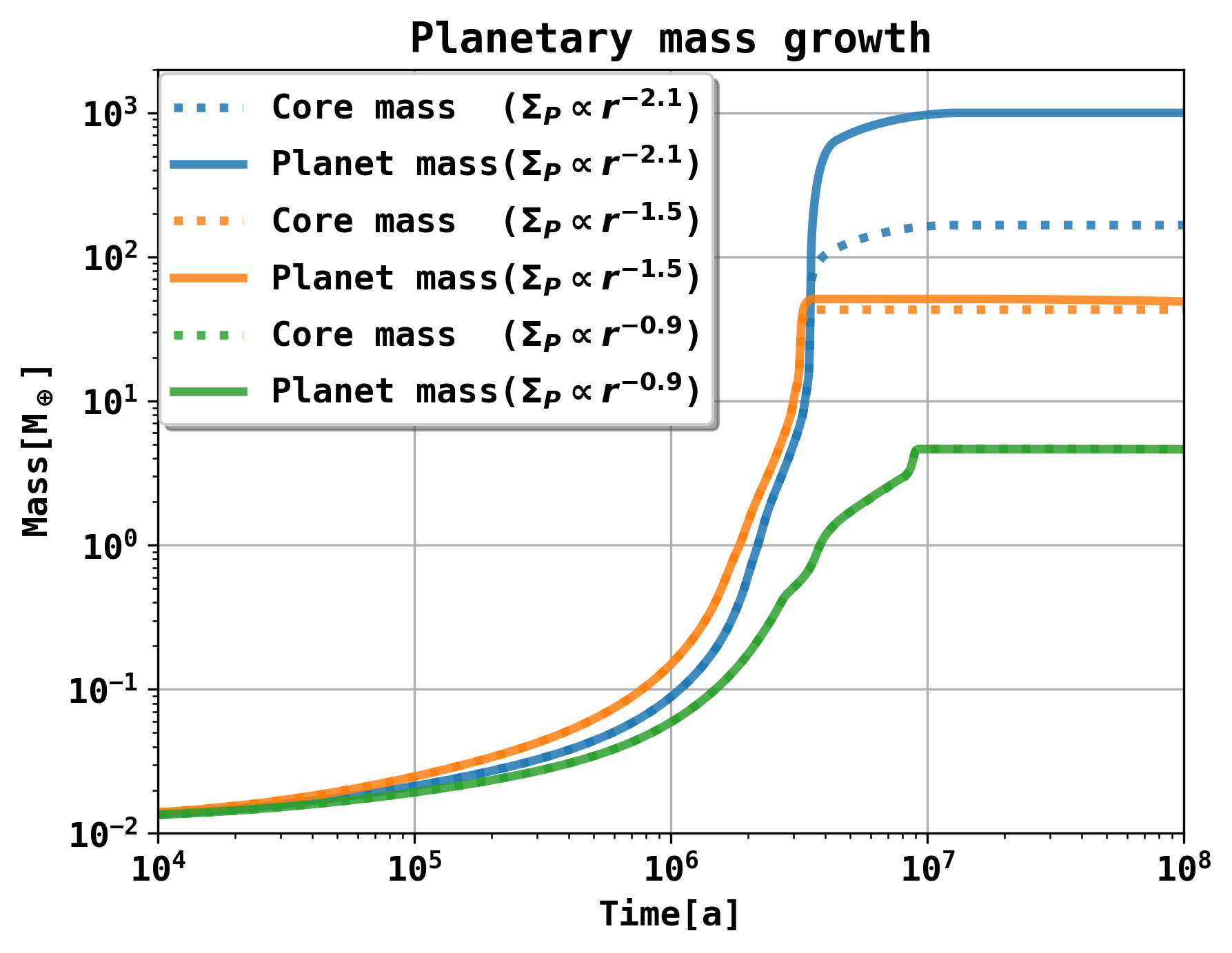}
\end{minipage}
\begin{minipage}{0.49\textwidth}
\centering
\includegraphics[width=\textwidth]{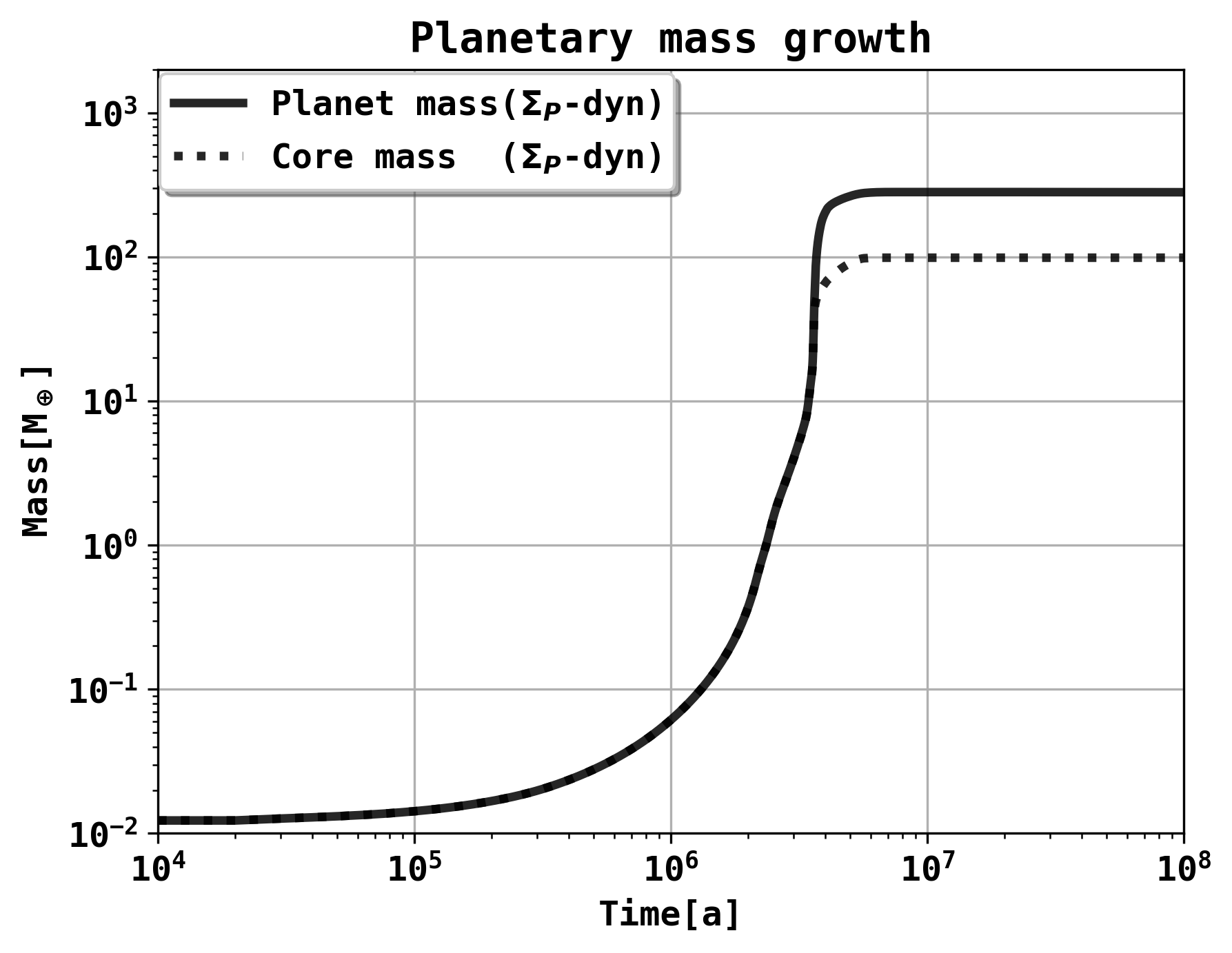}
\end{minipage}
\begin{minipage}{0.49\textwidth}
\centering
\includegraphics[width=\textwidth]{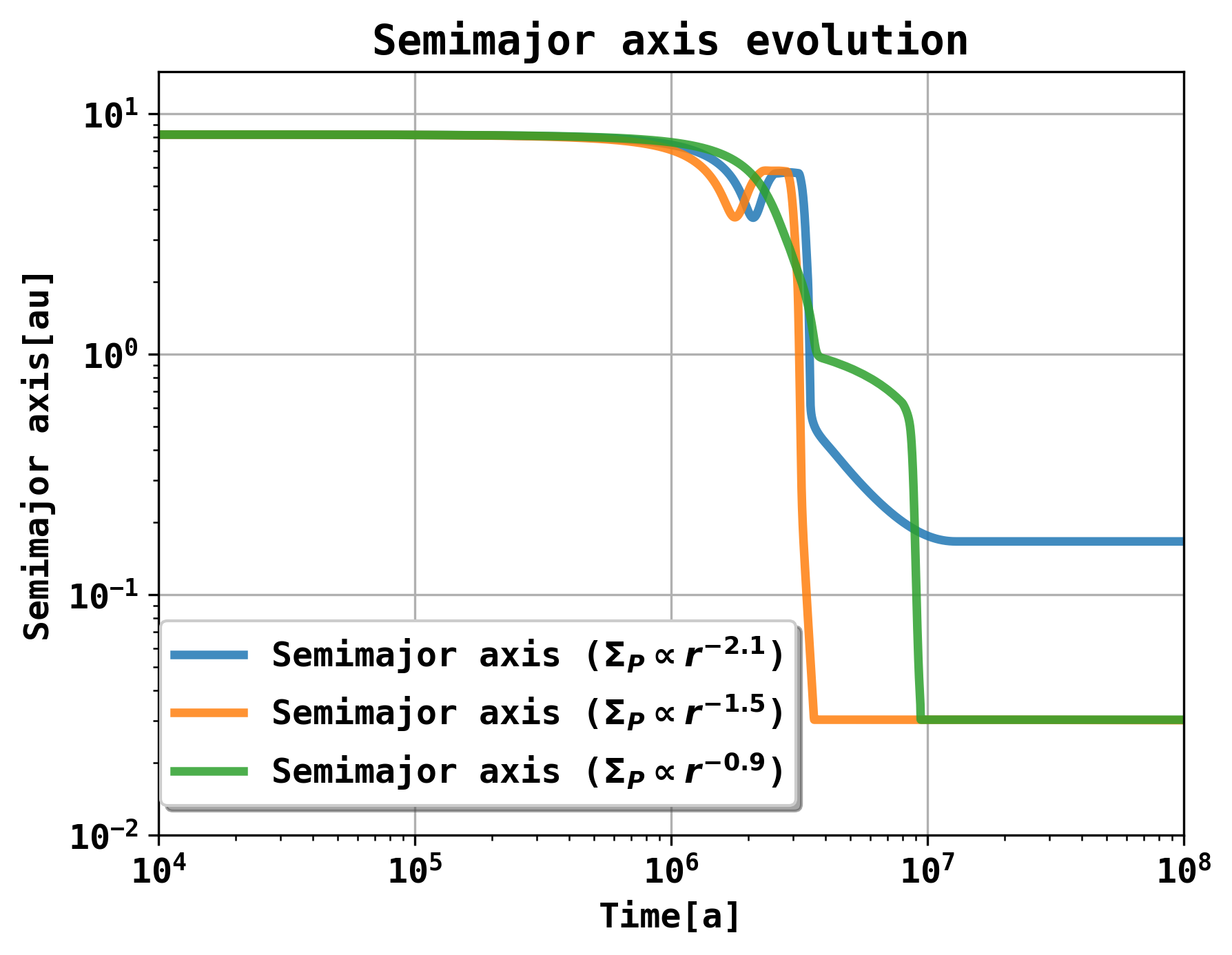}
\end{minipage}
\begin{minipage}{0.49\textwidth}
\centering
\includegraphics[width=\textwidth]{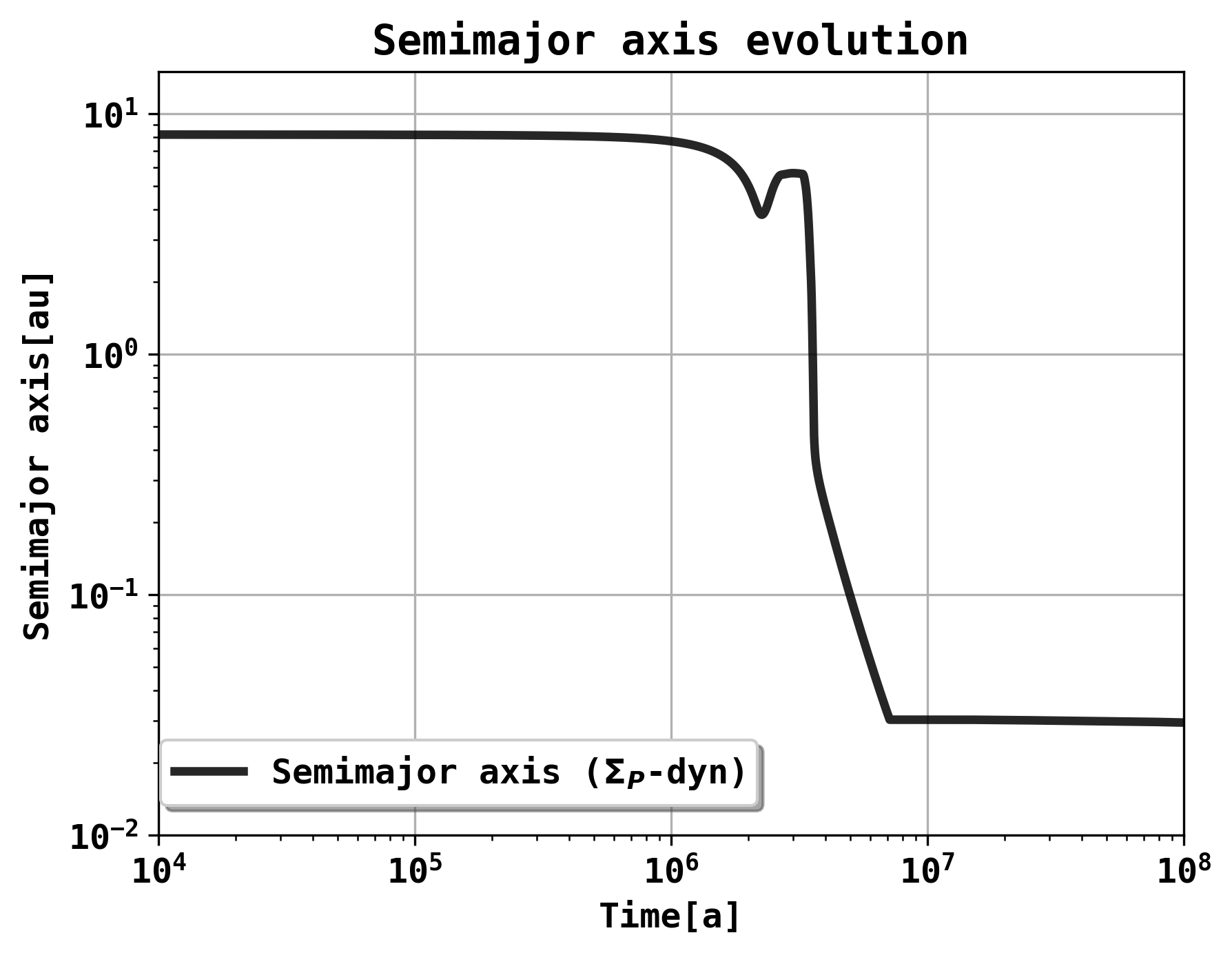}
\end{minipage}
\caption{Planetary growth tracks, mass over time and semimajor axis evolution for a giant planet system. The system that is studied leads to a gas giant planet of 997.6$\,$M$_{\oplus}$ for the $\Sigma_P \propto r^{-2.1}$ density distribution and a 281.7$\,$M$_{\oplus}$ planet for the dynamic model. The other systems lead to 51.1$\,$M$_{\oplus}$ for $\Sigma_P \propto r^{-1.5}$ and 4.64$\,$M$_{\oplus}$ for $\Sigma_P \propto r^{-0.9}$. The upper panel shows the mass and semimajor axis change during the evolution of the system, while the middle panel shows the growth of the embryo over time. The lower panel shows the semimajor axis evolution over time. Over Gyr timescales the giant planet in the dynamical planetesimal formation and the $\Sigma_P \propto r^{-1.5}$ simulation falls into the star due to tidal forces, which is no longer shown in the lower right panel. } 
\label{fig:Tracks}
\end{figure*}

Here we focus on a system that forms a 997.6$\,$M $_{\oplus}$ mass planet for the $\Sigma_P \propto r^{-2.1}$ density distribution and a 281.7$\,$M$_{\oplus}$ planet for the dynamical planetesimal formation run. The initial disk parameters for the setup are given in Table ~\ref{tab:initial_parameters}. Fig ~\ref{fig:Tracks} shows planetary growth tracks, the mass growth over time and the corresponding semimajor axis evolution.
The embryo in these systems was placed initially at 8.2$\,$au which seems to be a preferential starting location for giant planets, see Fig. ~\ref{fig:Starting_location}.
We can see that the higher planetesimal surface density has a drastic impact on the early stages of planetary growth. The planets in the $\Sigma_P \propto r^{-2.1}$ setup and the dynamical planetesimal formation run can grow fast enough to undergo runaway gas accretion, whereas the planet in  system $\Sigma_P \propto r^{-1.5}$ fails to do so, even though its core reaches a core mass of 43 M$_\oplus$. The planet in system $\Sigma_P \propto r^{-0.9}$ fails to build a large enough core for significant gas accretion and ends up at 4,64 M$_{\oplus}$.
\section{Discussion}
\label{Sec:Discussion}
\hubert{In our models with a fixed initial density slope for the planetesimals} we find that we can not form gas giant planets from planetesimal accretion with 100$\,$km sized planetesimals, if we assume that the surface density distribution of the planetesimals is shallow, varying as $r^{-0.9}$.
This is in agreement with studies from \cite{johansen2019exploring}, in which planetesimal accretion of large planetesimals is an inefficient accretion mechanism for low mass planetary embryos. Yet on the other hand we can clearly show that a change in the planetesimal surface density slope has a drastic effect on the global evolution of planetary systems. A steeper profile in the initial planetesimal surface density distribution can lead to gas giant growth in the inner region of protoplanetary disks, using only 100$\,$km sized planetesimals, while also forming a large amount of terrestrial planets and super earths. This result indicates that planetesimal accretion alone \hubert{can be} a very effective mechanism for planetary growth in the inner regions of circumstellar disks and can explain large diversities in the population of planets. 

\hubert{But more importantly}
we find that pebble flux regulated planetesimal formation leads \hubert{automatically from a shallow distribution of dust} to \hubert{a steep planetesimal distribution, leading to} much \hubert{higher planetary masses than in} the $\Sigma_P \propto r^{-1.5}$ density profile.
 
 The largest planets still can be formed using the $\Sigma_P \propto r^{-2.1}$ density slope and reach 1062.8 M$_\oplus$. The \hubert{most massive} planet in the $\Sigma_P \propto r^{-1.5}$ run reaches only 166.2 M$_{\oplus}$ and 7.8 M$_{\oplus}$ for $\Sigma_P \propto r^{-0.9}$. The maximum planetary mass for our dynamic simulation peaks at 317.1 M$_{\oplus}$. Comparing the mmsn ($\Sigma_P \propto r^{-1.5}$) profile with the dynamic formation model, we find that we increase the number of planets above 10$\,$M$_\oplus$ by 89$\%$ (from 159 to 301) and the number of planets above 20$\,$M$_\oplus$ by 345$\%$ (from 31 to 138), if we choose the formation of planetesimals to be consistent with the disks evolution.
 
 One has to keep in mind that the total mass in planetesimals is the lowest for the \hubert{dynamical} planetesimal formation model, since only a fraction of the dust and pebbles is transformed into planetesimals. The slope of the planetesimals that form over time however is steeper than the $r^{-1.5}$ slope. The total mass that is available for accretion is therefore lower in the planetesimal formation run, since pebble accretion on protoplanets is currently neglected.
 
We also find that \hubert{our current models} assuming 100$\,$km sized planetesimals do not form cold giants around the water iceline in any scenario due to orbital migration, although giant planets migrate trough that area, as a study of the initial embryo location in Fig. ~\ref{fig:Starting_location} shows. This might also indicate that the formation of planetesimals could \hubert{be enhanced} 
by the mechanisms around the iceline, as already predicted by \cite{Dr_kowska_2017} and \cite{schoonenberg2017planetesimal}. \\
They suggest, that sublimation and recondensation of icy pebbles at the iceline can have a drastic effect on the formation of planetesimals. This effect on planetesimal formation can be incorporated with our implemenatation by locally adapting the formation efficiency $\epsilon$ and promises to have a significant impact on the formation of heavy planets around the iceline. Finally, we find that the placement of planetary embryos appears to be a strong component for giant planet formation, see Fig. ~\ref{fig:Starting_location}. The effect of the starting location of planetary embryos in combination with the formation of planetesimals can be studied in future work in greater detail, including the dynamical placement of planetary seeds during the evolution of the disk. In combination with the increased planetesimal formation around the iceline and pebble accretion, we believe these features to have a drastic impact on our synthetic planet populations. We expect this to explain the abundance of cold/hot giants and terrestrial planet diversity.
\section{Summary $\&$ Outlook} 
\label{summary}
Using the two population solid evolution model by \cite{birnstiel2012simple} and the pebble flux regulated model for planetesimal formation by \cite{Lenz_2019}, we have studied the effect of planetesimal formation using our model for planetary population synthesis.
Comparing the \hubert{dynamical} planetesimal formation with different \hubert{ad-hoc} planetesimal surface density distributions, we find strong differences for the formation of planets in the inner parts of cirscumstellar disks for a planetesimal size of 100$\,$km. This can be linked directly to the steeper slope in $\Sigma_P$ as reference simulations with shallower surface density profiles show. We hereby show the impact of the planetesimal surface density distribution and formation on the population of planets.
The main results of planetesial formation for single embryo planet population synthesis are:
\\
\noindent
\begin{itemize}
    \item Planetesimal accretion with 100$\,$km sized planetesimals can be a very efficient planetary growth mechanism in the inner regions of circumstellar disks and creates a large variety of planets.\\
    \item Pebble flux regulated planetesimal formation enables gas giant formation by accreting only 100$\,$km sized planetesimals, due to highly condensed planetesimal areas in the inner regions of circumstellar disks.\\
    \item Pebble flux regulated planetesimal formation fails to form cold giant planets outside the iceline. The reason however is not a too long core accretion timescale compared to the disk lifetimes, but orbital migration that removes the cores faster than they can grow.\\
    \item We no longer rely on an ad hoc assumption like the mmsn model for the distribution of planetesimals in protoplanetary disks\hubert{, but can start with much shallower mass distributions in agreement with observations of disks around young stars.} \\
    \item Dynamic planetesimal formation increases the amount of planets above 10$\,$M$_E$ by 89$\%$ and the number of planets above 20$\,$M$_E$ by 345$\%$ compared to the mmsn hypothesis.
\end{itemize}
The biggest technical advantages that the newly implemented solid evolution model brings are:
\\
\noindent
\begin{itemize}
\item Pebble accretion can be included next to planetesimal accretion into our population synthesis framework to study their individual contributions to planetary growth. \\
\item Locally adapting the planetesimal formation efficiency $\epsilon$ gives us the opportunity to study increased planetesimal formation around the iceline, or other dynamically evolving planetesimal surface density profiles, like e.g. rings in disks. \\
\item Planetary embryo formation based on the local planetesimal surface density evolution can be incorporated.
\end{itemize}
These improvements will enable us to consistently study the full size range of planet formation in a globally coupled framework, beginning from a disk of gas and dust.

\begin{acknowledgements}
This research is supported by the German Science Foundation (DFG) under the priority program SPP 1992: "Exoplanet Diversity" under contract KL 1469/17-1, by the priority program SPP 1385 "The first ten million years of the Solar System" under contract KL 1469/4-(1-3) "Gravoturbulent planetesimal formation in the early solar system" and SPP 1833 "Building a Habitable Earth" under contract KL 1469/13-(1-2) "Der Ursprung des Baumaterials der Erde: Woher stammen die Planetesimale und die Pebbles? Numerische Modellierung der Akkretionsphase der Erde." and DFG Research Unit FOR2544 “Blue Planets around Red Stars” under contract KL 1469/15-1. This research was also supported by the Munich Institute for Astro- and Particle Physics (MIAPP) of the DFG cluster of excellence "Origin and Structure of the Universe" and was performed in part at KITP Santa Barbara by the National Science Foundation under Grant No. NSF PHY11-25915. 
C.M. and A.E. acknowledge the support from the Swiss National Science Foundation under grant BSSGIO\_155816 ``PlanetsInTime''. The authors thank Martin Schlecker and Remo Burn for many fruitful discussions.
\end{acknowledgements}
%%%%%%%%%%%%%%%%%%%%%%%%%%%%
% for the bibliography, at the end
\bibliography{Template.bib} % your references Yourfile.bib
%%%%%%%%%%%%%%%%%%%%%%%%%%%%
%%%%%%%appendix
%\appendix
%\section{Disk masses}
%\begin{appendix}
%%%%%%%%Disk masses
%%%%%%%%%%%%%%
%%%%%%%%%%%%%% Planet masses
%\section{Planet masses}
%\section{Planet masses}
%%%%%%%%% Mass occurences - 1000au
%%%%%%%%%% Mass occurences < 1au
%%%%%%%%%%%%%%
%%%%%%%%%%%%%% Gas giant growth
%\section{Gas giant growth}
%%%%%%%%%Table
\begin{table*}
    \caption{Initial parameters for the simulation from Fig. ~\ref{fig:disk_evolution} that leads to a 997.6 M$_\oplus$ gas giant planet for the $\Sigma_P \propto r^{-2.1}$ distribution and a 281.7 M$_{\oplus}$ gas giant planet in the dynamical planetesimal formation run. The parameters for the planetesimal formation are the same for every simulation.}
    \centering
    \begin{tabular}{l l l}
    \hline
    \hline 
    Symbol                  &  Value                          & Meaning                           \\
    \hline    \\
    M$_\text{disk}$         &  0.128$\,$M$_\odot$             & Total mass of the gas disk        \\
                  
    a$_\text{in}$           &  0.03$\,$au                     & Inner planetesimal disk radius    \\
      
    a$_\text{out}$          &  137$\,$au                      & Exponential cutoff radius         \\
                  
    $d_g$                   &  3.2 $\times 10^{-2}$           & Dust to gas ratio                 \\
                                        
    $\alpha$                &  1.0 $\times 10^{-3}$          & Turbulence parameter              \\
                                        
    a$_\text{start}$        &  8.2$\,$au                     & Embryo starting location          \\
                                        
    M$_\text{emb}$          &  0.0123$\,$M$_{\oplus}$          & Embryo starting mass              \\
                                        
    M$_\text{wind}$         &  2.87$\times 10^{-5}$ $\,$ M$_{\odot}$/year   & Photoevaporation rate             \\
    
    \hline                            \\
    Planetesimal formation parameters \\
    \hline                            \\
                                            
    v$_\text{frag}$         &  10$\,$m/s                      & Fragmentation velocity (pebbles)  \\
                                            
    $\epsilon/d$            &  0.01                           & Planetesimal formation efficiency \\
                                            
    $\rho_s$                &  1.0 g/cm$^{3}$                 & Planetesimal solid density        \\
                                              
    $\text{St}_\text{min}$  &  0                              & Min Stokes number for planetesimal formation \\
                                                  
    $\text{St}_\text{max}$  &  $\infty$                       & Max Stokes number for planetesimal formation  \\
    \hline                       
    \\
    \end{tabular}
    \label{tab:initial_parameters}
\end{table*}
%%%%%%%%%Growth tracks
%\end{appendix}
\end{document}